\documentclass[12pt]{article}
\usepackage[sort,longnamesfirst]{natbib}

\usepackage{amsbsy,amsmath,amsthm,amssymb,mhchem,graphicx}
\usepackage[usenames,dvipsnames]{color}
\usepackage{subfigure}
  \usepackage{setspace}
  \usepackage{wrapfig}
\usepackage{geometry}
\newcommand{\bdm}{\begin{displaymath}}
\newcommand{\edm}{\end{displaymath}}
\newcommand{\beq}{\begin{equation}}
\newcommand{\eeq}{\end{equation}}

\newcommand{\eps}{\epsilon}

\newcommand{\bbeta}{ \mbox{\boldmath $\beta$}}

\newcommand{\btheta}{ \mbox{\boldmath $\theta$}}

\newcommand{\bepsilon}{ \mbox{\boldmath $\epsilon$}}
\newcommand{\bdelta}{ \mbox{\boldmath $\delta$}}

\newcommand{\bzeta}{ \mbox{\boldmath $\zeta$}}
\newcommand{\boldeta}{ \mbox{\boldmath $\eta$}}

\newcommand{\btau}{ \mbox{\boldmath $\tau$}}

\newcommand{\bY}{\mathbf{Y}}

\newcommand{\bt}{\mathbf{t}}

\newcommand{\bx}{\mathbf{x}}
\newcommand{\by}{\mathbf{y}}

\newcommand{\bI}{\mathbf{I}}

\newcommand{\cL}{{\cal L}}

\begin{document}

\begin{singlespace}
\title{
\vspace{-1in}
Upscaling Uncertainty with Dynamic Discrepancy for a Multi-scale Carbon Capture System 
}
\author{K. Sham Bhat, David S. Mebane, Curtis B. Storlie, and Priyadarshi Mahapatra
}
\date{}

\maketitle
\begin{abstract}
 
Uncertainties from model parameters and model discrepancy from small-scale models impact the accuracy and reliability of predictions of large-scale systems. Inadequate representation of these uncertainties may result in inaccurate and overconfident predictions during scale-up to larger models. Hence multiscale modeling efforts must quantify the effect of the propagation of uncertainties during upscaling. Using a Bayesian approach, we calibrate a small-scale solid sorbent model to Thermogravimetric (TGA) data on a functional profile using chemistry-based priors. Crucial to this effort is the representation of model discrepancy, which uses a Bayesian Smoothing Splines (BSS-ANOVA) framework. We use an intrusive uncertainty quantification (UQ) approach by including the discrepancy function within the chemical rate expressions; resulting in a set of stochastic differential equations. Such an approach allows for easily propagating uncertainty by propagating the joint model parameter and discrepancy posterior into the larger-scale system of rate expressions. The broad UQ framework presented here may have far-reaching impact into virtually all areas of science where multiscale modeling is used. 
	 
\end{abstract}
\textbf{Short Title: Upscaling Uncertainty for a Carbon Capture System} \\
\end{singlespace}
\vspace{-0.3in}
\section{Introduction}\label{sec:intro}

The Carbon Capture Simulation Initiative (CCSI) sponsored by the U.S. Department of Energy is focused on accelerating the
adoption of new carbon capture technology using modeling and simulation to reduce the amount of physical testing required for development of larger-scale power plants. This effort increases the reliance on computer models for upscaling.  These carbon capture systems involve phenomena at the quantum scale up through to the industrial macroscale which are analyzed using complex computer models.%; three scales of the multi-scale carbon capture system are shown in Figure \ref{fig:scaleup}.    
%\begin{wrapfigure}{r}{.5\textwidth}
%\vspace{-.45in}
%\begin{center}
%\caption{Overview of the scales in a multi-scale carbon capture system.}
%\vspace{-.10in}
%\includegraphics[width=0.5\textwidth]{Figures/upscaling1.png}
%%  \figoneAC{Figures/fullupscdiag.pdf}
%\vspace{-0.0in}
%\label{fig:scaleup}
%\end{center}
%\vspace{-.85in}
%\end{wrapfigure} 
 
\begin{figure}
\vspace{-.65in}
\begin{center}
\caption{\textit{Functional input (temperature) in green, functional output (sorbent weight gain) in blue.}}
\vspace{-0.15in}
\includegraphics[width=0.8\textwidth]{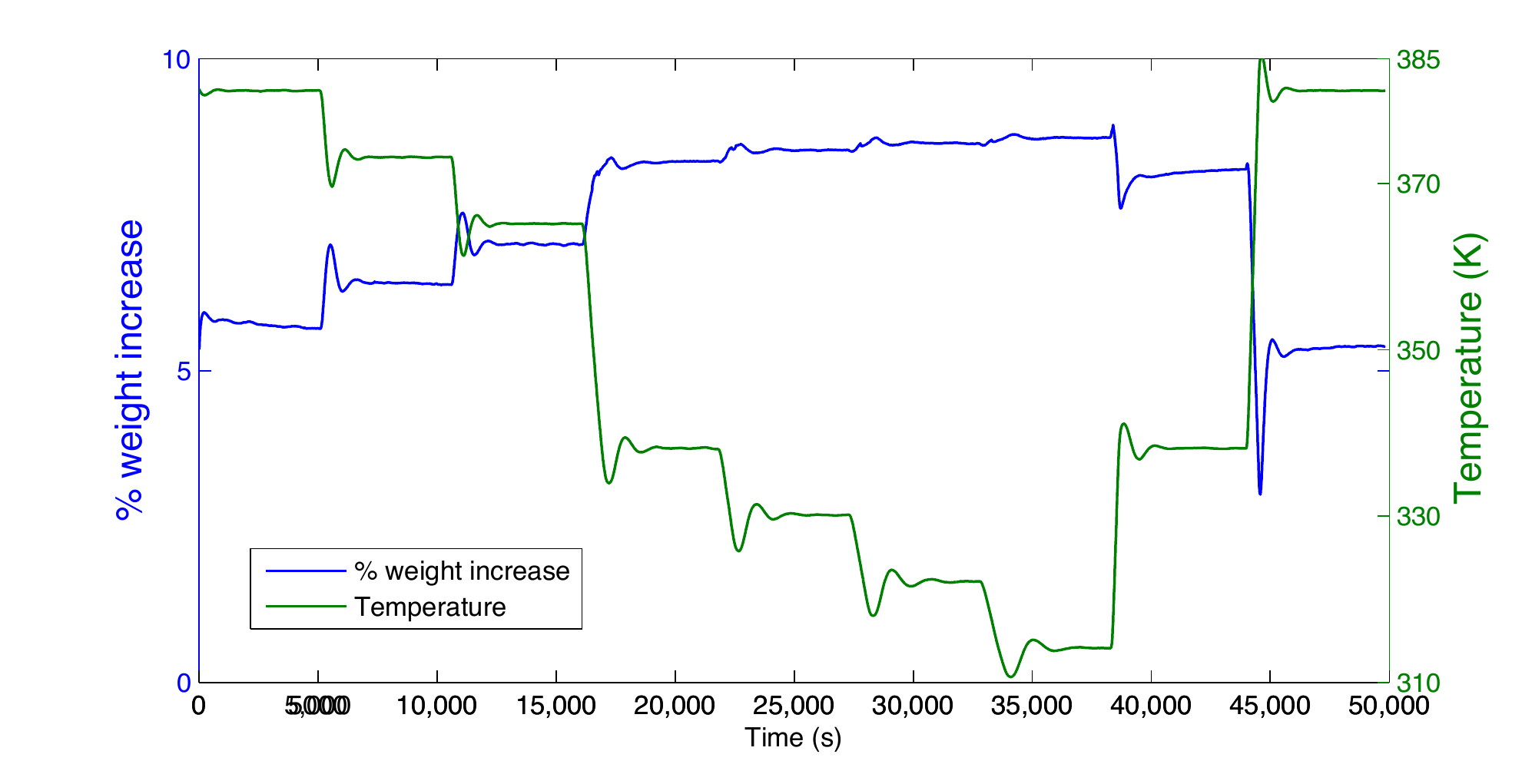} 
%  \figoneAC{Figures/fullupscdiag.pdf}
\vspace{-0.0in}
\label{fig:funcform}
\end{center}
\vspace{-.35in}
%\end{wrapfigure}  
\end{figure}

The multi-scale uncertainty quantification effort in this article is illustrated on a simple carbon capture process for a ``bubbling fluidized bed" absorber \citep{lee2012one}, which is built using Aspen Custom Modeler \citep{aspen2011version}.  The major driver of the uncertainty in the system is the chemical sorbent model, characterized by one (or more) chemical reactions, dependent on several chemical parameters which describe the equilibrium and kinetic facets of the reactions.  In addition, there are certain \textit{system conditions} (or physical inputs e.g. temperature and partial pressure) that affect the behavior of the system. 
The small-scale model takes in temperature and pressure inputs (describing possible system conditions) and a set of chemistry model parameters and outputs the sorbent weight gain; experimental data are collected using Thermogravimetric Analysis (TGA) with the same inputs and outputs.  In Figure \ref{fig:funcform}, an input (temperature) and output (sorbent weight gain) from the small-scale sorbent model are shown; note that both the inputs and outputs to the model are functional (in time).      
Both the small-scale sorbent model, isolated to only a chemical reaction with no fluid dynamics (small-scale), and the fully coupled model for a process system (large-scale) are governed by the solution of one or more rate-based differential equations.    

%\begin{wrapfigure}{r}{.7\textwidth}

Complex computer models are widely used by scientists and engineers to understand and predict the behavior of complex physical processes (e.g., climate change, nuclear reactor performance, fluid transport, and carbon capture systems) when direct experimentation is difficult, expensive, or impossible.  While these computer models are often grounded in scientific theory, they will still have various forms of uncertainty that must be accounted for when used to predict the behavior of the physical process.  These uncertainties may come from many sources; for example, incomplete information about physical constants and/or inadequate quantitative models to describe the physical behavior.  Hence, uncertainty quantification (UQ) is widely recognized as essential to analysis of complex computer models. \citep{kennedy:bcc,Oakley04,higdon2008cmc,Storlie09b}.
Many of these uncertainties arise in a multi-scale system context, and uncertainties at a small scale may greatly impact the accuracy and reliability of large system predictions.  Therefore, it is crucial that this small-scale uncertainty be appropriately represented.  
%It is crucial that this small-scale uncertainty be appropriately accounted for, as inadequate representation of uncertainty may result in inaccurate and overconfident predictions when models are scaled up to make predictions of large scale systems.  
%Hence it should be among the highest priorities for a designer of a large scale system to determine and quantify the effect of the uncertainties when upscaling from a small model to the large-scale system.  It should also be noted that quantifying uncertainties directly at the large scale is often not feasible due to the scarcity of data. \\

Two major sources of uncertainty in predictions from computer models are parameter uncertainty and model form discrepancy; statistical calibration of the model to experimental (or field) data provides a means to formally quantify these uncertainties.  \cite{kennedy:bcc} developed a framework for calibration of computer models, which includes a model discrepancy term describing the deviation from the model to reality.  Failure to properly account for model discrepancy can lead to overfitting of the model parameters \citep{bayarri2007fvc}, and hence inaccurate predictions.   Constructing the model form discrepancy using expert scientific knowledge about how the model does not to conform to reality will improve inference and resulting model predictions \citep{brynjarsdottir2012} as well as confounding issues between the parameters and discrepancy \citep{liu2009bay}.   Furthermore, such an approach provides a theoretical or empirical representation of how the model is flawed, which could be useful for how to improve the model.  In this paper, we propose an efficient means to propagate forward the uncertainty in model form discrepancy, along with parameter uncertainty, resulting from model calibration at small scale, when making predictions at scale.
%Furthermore, such an approach provides a theoretical or empirical representation of how the model is flawed, which could possibly be used to justify some extrapolation during upscaling to larger scale systems.    

%Upscaling uncertainty for a multi-scale system presents two major challenges: (i) forward propagation of the full uncertainty including the model form discrepancy, (ii) potentially a large amount of uncertainty during upscaling due to large functional input space for the large-scale system.  
Upscaling uncertainty for a multi-scale system via forward propagation of both the model parameter and discrepancy sources of uncertainty presents a major challenge.  Many current approaches for forward uncertainty propagation to a new system (e.g. large scale) include only the uncertainty from the calibrated model parameters. This may be problematic since the calibrated model output, even at the ``true" parameters, may not be an accurate representation of the system as it ignores the effect of model form discrepancy.  The typical approach to include model discrepancy is to use a fully non-intrusive ``black-box" approach for calibration, i.e. obtain a joint distribution of the model parameters and discrepancy, and propagate that distribution to the large-scale system.  However, this approach is not always feasible in practice.  The estimated model discrepancy at small-scale may not be relevant to that at large scale.  For instance, often the inputs and even the quantity of interest are not comparable from the calibration data to prediction.
%However that approach is quite troublesome in practice, particularly when there are functional inputs, often requiring considerable effort to propagate posterior discrepancy realizations to the large-scale model.  
%In addition, the input space of the large-scale system is generally very large, there are a large number of paths that the system can possibly take, and hence it is difficult to replicate the unknown system conditions in bench-scale experiments.  Hence, data obtained at the small-scale level is often very sparse in functional space, and upscaling invariably results in extrapolation.  The uncertainty due to this extrapolation is inadequately represented when model form discrepancy is not upscaled.        

We propose a novel approach to describe the dynamic discrepancy between the small-scale model and reality, which has a clear scientific interpretation about the model failings and is efficient for both calibration and upscaling.  Our approach uses an intrusive uncertainty quantification (UQ) approach by including the stochastic discrepancy function within the small-scale sorbent model equations, resulting in a stochastic differential equation(s) (SDE).   In addition, the discrepancy terms may interpreted as a "correction" for a specific physical or chemical deficiency in the small-scale sorbent model (e.g. deviation from the ideal gas assumption).   Model form discrepancy here is represented using a Gaussian Process with a Bayesian Smoothing Spline (BSS)-ANOVA covariance.  The BSS-ANOVA framework has many advantages within this context; it provides an approximate parametric form which is very convenient for both calibration and upscaling, accounts for the uncertainty due to extrapolation while upscaling, and provides substantial computational gains requiring only $\mathcal{O}(N)$ computational time by eschewing matrix inversions.   

Using a Bayesian approach, we calibrate the small-scale sorbent model to data; resulting in a joint sample-based distribution of both model parameters and discrepancy basis function coefficients. The uncertainty in model parameters \textit{and} discrepancy are then easily propagated to scale by passing these posterior samples into the large-scale system of rate expressions and solving them in a Monte Carlo fashion.  The output from the large-scale system provides both a quantity of interest (such as the proportion of CO$_2$ captured) as well as a functional profile of the system conditions (which may also be seen as inputs to the small-scale sorbent model).  
To the best of our knowledge, there is not an uncertainty quantification approach which simultaneously incorporates a stochastic discrepancy in a calibration framework with functional inputs and forward propagates both parameter and model form uncertainty across scales in an efficient manner.

%This approach not only addresses the incorporation of uncertainty but also facilitates the process of model development as sorbent models of varying complexity are entertained, ranging from a relatively simple lumped kinetic model that combines the behavior of multiple kinetic effects to more complex models that explicitly account for more detailed chemical behavior.  Our approach allows for forward propagation of the different sources of uncertainty present in the first principles sorbent model to the larger scale process system. 
%The results from this model are used as input to a process model, incorporating known sources of uncertainty including model discrepancy.  Subsequent results provide information on behavior of larger scale systems, extrapolating beyond the scale of existing physical systems, while incorporating information about parameter uncertainty, model discrepancy, and data uncertainty.

The rest of this paper is organized as follows.   In Section \ref{sec:multscalemodel}, we give a broad overview of our framework to upscale uncertainty with dynamic discrepancy, and we discuss the details of our calibration, dynamic discrepancy, and upscaling approach in Section \ref{sec:calibdyndisc}.  We demonstrate our methodology on a small carbon capture system with a chemical kinetics sorbent model in Section \ref{sec:toymodelapp}.  Finally, we conclude with a discussion, caveats, and avenues for future research in Section \ref{sec:discussion}.

\section{Uncertainty Quantification for a Multi-scale System}\label{sec:multscalemodel}

This section contains a broad overview of our approach to upscaling uncertainty in a multi-scale system.  The goal is to quantify uncertainty of a large-scale system, with many small-scale physical processes embedded in it; where the behavior of both the large-scale system and the small-scale processes are simulated with deterministic computer models.  In this paper a single small-scale model is used for the ease of demonstration.  The small-scale sorbent model and the large-scale carbon capture process system are discussed in Section \ref{subsec:basicchem} and \ref{subsec:largescaleeq}, respectively.  In Section \ref{subsec:upsunc}, we provide an overview of the proposed Bayesian dynamic discrepancy approach used for UQ analysis, which has several advantages in this particular framework; incorporation of functional inputs, convenience in propagating information across scales, and the inclusion of a dynamic discrepancy function to describe shortcomings of the model.

\subsection{Basic Chemical Kinetics Model}  \label{subsec:basicchem}
%A small-scale chemical kinetic sorbent model is introduced here as a sandbox for demonstrating our discrepancy and upscaling methodology.   
The basic chemistry model describes the adsorption of CO$_2$ by a solid Amine based sorbent. It was determined in the course of an {\em ab initio} study documented in reference \citep{mebane2013bayesian} that this adsorption of CO$_2$ takes place through the formation of carbamic acid according to the reaction:
  \begin{equation}
    \label{eq:carbacid}
    2\textrm{R}_2\textrm{NH} + \textrm{CO}_2 (\textrm{g}) \rightleftharpoons \textrm{R}_2\textrm{NCOOH}:\textrm{R}_2\textrm{NH}
  \end{equation}
  
  \begin{table}[t!]
\caption{Summary of Inputs, Outputs, and small-scale sorbent parameters.}
\vspace{-0in}
\framebox{
\small
\begin{minipage}{1\textwidth}
\begin{itemize}
 \setlength{\itemindent}{-.275in}
 \setlength{\leftmargin}{-.275in}
\item[] $\!\!\!\!\!\!\!$Experimental Outputs and State Variables (a function of time) $\by$:\\[-.3in]
 \begin{itemize}
 \setlength{\itemindent}{-.3in}
 \setlength{\leftmargin}{-.3in}
 \item[$\by$]: Weight fraction of sorbent. $w$.
 \end{itemize}
\item[] $\!\!\!\!\!\!\!$Functional Input Profile (potential system conditions) $\bzeta(t)$:\\[-.3in]
 \begin{itemize}
 \setlength{\itemindent}{-.3in}
 \setlength{\leftmargin}{-.3in}
 \item[$\bzeta_1$]: Temperature $T \in [310, 380]$ K
 \item[$\bzeta_2$]: Partial pressure of $CO_2, ~ p \in [0, 100]$\%
 \end{itemize}
\item[] $\!\!\!\!\!\!\!$Sorbent Model Parameters $\btheta$:\\[-.3in]
 \begin{itemize}
 \setlength{\itemindent}{-.45in}
 \setlength{\leftmargin}{-.45in}
  \item[$\theta_1$]: Reaction Enthaply  $\Delta H \in [-120,-30]$ kJ/mol
 \item[$\theta_2$]: Reaction Entropy $\Delta S \in [-450, -200]$ J/mol-K
 \item[$\theta_3$]: Activation Energy (kinetic) $\Delta H^\ddag \in [-150, -50]$  kJ/mol
 \item[$\theta_4$]: Kinetic entropy plus other parameters $\gamma \in [0,10]$ 
 \item[$\theta_5$]: Amine site density $n_v \in [1000, 2351]$ mol/m$^3$
 \end{itemize}
\end{itemize}
\end{minipage}
}
\label{tab:data_descr}
\vspace{.1in}
\end{table}
  
 A summary of the inputs and outputs may be found in Table \ref{tab:data_descr}.  There are five model parameters to be estimated: $\btheta=[\Delta H, \Delta S,\Delta H^\ddag,\gamma,n_\textrm{v}]$. The equilibrium parameters are $\Delta H$, $\Delta S$ and $n_\textrm{v}$ and the kinetic parameters are $\Delta H^\ddag$ and $\gamma$.  The rate equations of this sorbent model in Equation (\ref{eq:kinetic}) are solved on a temporal grid; potential system conditions, temperature ($T$) and partial pressure ($p$) of CO$_2$, are functional inputs over time (see Figure \ref{fig:funcform} for an example temperature input), resulting in a functional response $w(t)$ (the weight increase of the sorbent). 
  \begin{gather}
    \label{eq:kinetic}
    \frac{\partial x}{\partial t} = k\left[(1-2x)^2p - x^2/\kappa\right]\\
    \notag
    w = Mn_\textrm{v}x/\rho\\
    \notag
    \kappa^E = \exp{(\Delta S/R)}\exp{(-\Delta H/RT)}/P\\
    \notag
    \kappa^K = \gamma T\exp{(-\Delta H^\ddag/RT)}
  \end{gather}  

The response $w(t)$ is a multiple of the chemical state $x$, or the fraction of amine sites occupied by carbamic acid. Additional constants within the model are $M$ is the molar weight of CO$_2$, $\rho$ is the sorbent density, $R$ is the ideal gas constant, and $P$ is the total pressure.  The equilibrium constant $\kappa^E$ is a function of $\Delta S$ and $\Delta H$, while the reaction rate constant $\kappa^K$ is a function of $\gamma$ and $\Delta H^\ddag$.  As can be seen from Figure \ref{fig:funcform}, as temperature decreases there is an increase in sorbent uptake of CO$_2$ resulting is an increase in the weight of the sorbent.   Similarly, as pressure increases, there is an an increase in sorbent uptake of CO$_2$ and in the weight of the sorbent.  The experimental apparatus used to obtain the data here requires that the partial pressure remain constant over time, so the functional pressure input is a constant line in these data.  However, several time series observations are collected at different (constant over time) partial pressures.  
%  where $x$ is the fraction of amine sites occupied by carbamic acid;
%  $p$ is the partial pressure of CO$_2$ in the gas; $w$ is the
%  fractional weight gain recorded in the TGA; $M$ is the molar weight
%  of CO$_2$; $n_\textrm{v}$ is the number of active amine sites per
%  unit volume of sorbent; $\rho$ is the sorbent density; $\kappa$ is
%  the equilibrium constant, which is a function of the entropy $\Delta
%  S$ and enthalpy $\Delta H$ of the reaction, the temperature $T$ and
%  the total pressure $P$; $k$ is the reaction rate constant, which is
%  a function of the preexponential factor $\gamma$, the barrier
%  enthalpy $\Delta H^\ddag$ and the temperature; and $R$ is the ideal
%  gas constant.  
%More details about this sorbent model and the carbon capture process model are discussed in Section \ref{sec:toymodelapp} and in the Supplementary Material.
\subsection{Overview of Large-Scale Model}  \label{subsec:largescaleeq}
The coupled large-scale system is introduced here with a single quantity of interest $x$; this framework may be easily extended to the multivariate case.  The system conditions are $\bzeta(t)=[\zeta_1(t),\dots,\zeta_q(t)]$, where $\zeta_i(t)$ represents the curve for the $i$th condition at time $t$. The function space for these system conditions is very large as it must include any physically feasible curve.
%A set of small-scale model parameters $\btheta$ are the inputs into this large-scale system.    
\begin{wrapfigure}{l}{.5\textwidth}
\vspace{-.45in}
\begin{center}
\caption{Process model for simple carbon capture system.}
\vspace{-.00in}
\includegraphics[width=0.5\textwidth]{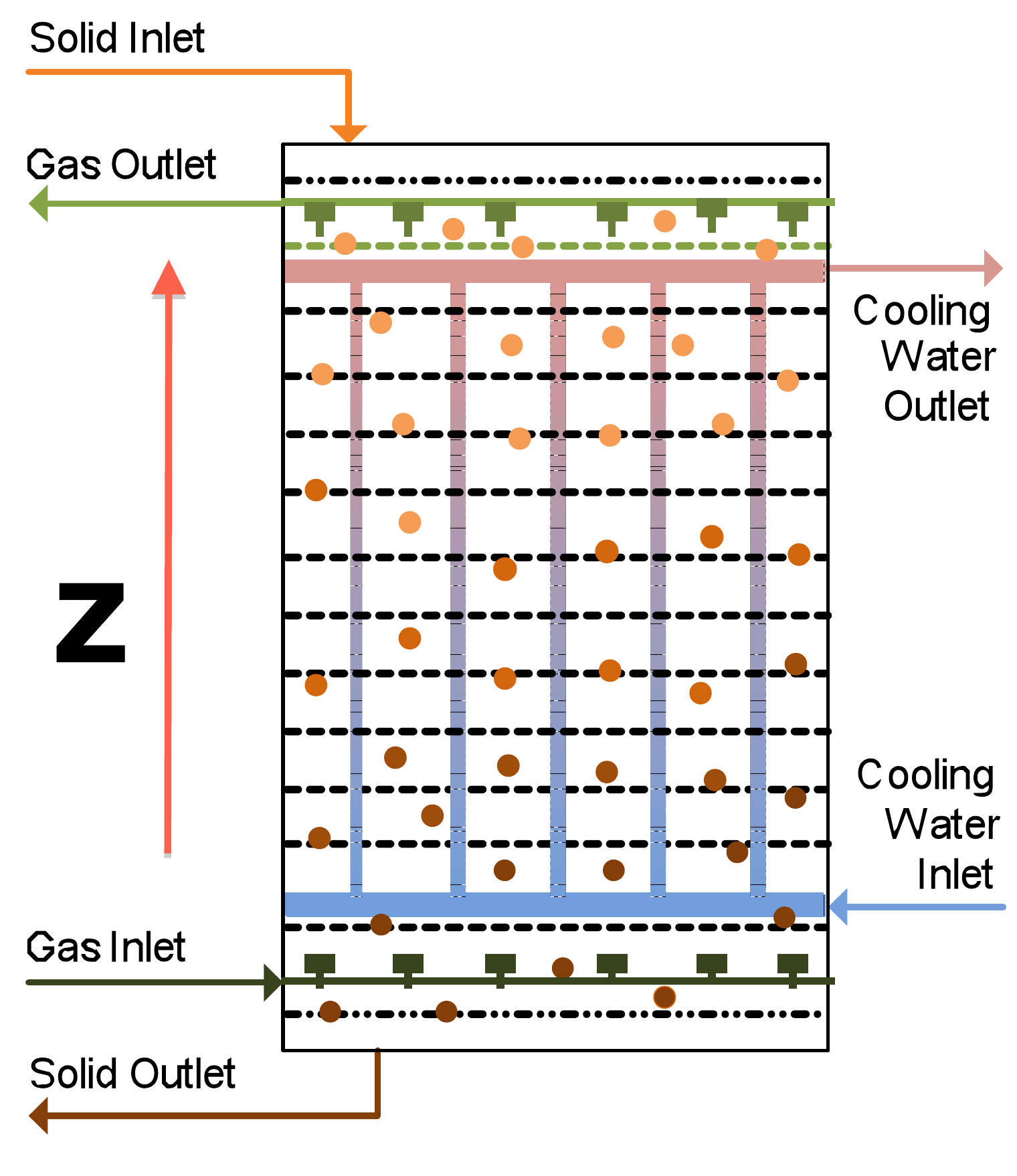} 
%  \figoneAC{Figures/fullupscdiag.pdf}
\vspace{-0.0in}
\label{fig:procmod}
\end{center}
\vspace{-2.05in}
\end{wrapfigure}  
A general rate equation describing the small-scale model is provided below in Equation (\ref{eq:smallmod}) for convenience, where $x$ represents a state variable and $f_s(x,\bzeta(t);\btheta)$ is the chemical rate function. 
 \begin{gather}
    \label{eq:smallmod}
    \frac{\partial x}{\partial t} = f_s(x,\bzeta(t);\btheta)
\end{gather}
A simple carbon capture process model is shown in Figure \ref{fig:procmod}. The large-scale system may be generally expressed as a set of differential equations below in Equations (\ref{eq:largesysx1})-(\ref{eq:largesysx2}); these equations are solved on a one-dimensional spatial grid (denoted by the variable $z$).  
 \begin{align}
    \label{eq:largesysx1}
   & \frac{\partial x}{\partial z}= f_s(x,\bzeta(t);\btheta) \\
  %  \frac{\partial \bzeta_1}{\partial t} &= g_1(x,\bzeta(t);\btheta)\\
  & g_1(x,\bzeta(z);\btheta)=0\\
    \nonumber
    &~~~~~~~~~\vdots \\
     \label{eq:largesysx2}
    %\frac{\partial \bzeta_q}{\partial t} &= g_q(x,\bzeta(t);\btheta)
  &  g_q(x,\bzeta(z);\btheta)=0;
\end{align}
The outputs from the large-scale system are functional curves (as a function of $z$) of the quantity of interest $x$ and the system conditions $\bzeta$.  For direct comparison with the inputs and outputs to the small-scale sorbent model, it is convenient to represent these functional curves as a function of time, which is accomplished using a conversion along the flow-velocity field.  The details of this conversion are provided in the Supplementary Material.   
%The details of resolving the spatial grid $z$ into a
%At each location on the spatial grid, the time of the system can be computed, and the resulting set of times may be considered as a temporal grid.         
%The differential equations of the process model are the Euler-Lagrange equations (add references) pertaining to the functional minimization.   
\subsection{Framework for Upscaling Uncertainty} \label{subsec:upsunc}
\begin{wrapfigure}{l}{.62\textwidth}
\vspace{-.65in}
\begin{center}
\caption{Overview of the upscaling process.}
\vspace{-0.0in}
\includegraphics[width=0.6\textwidth]{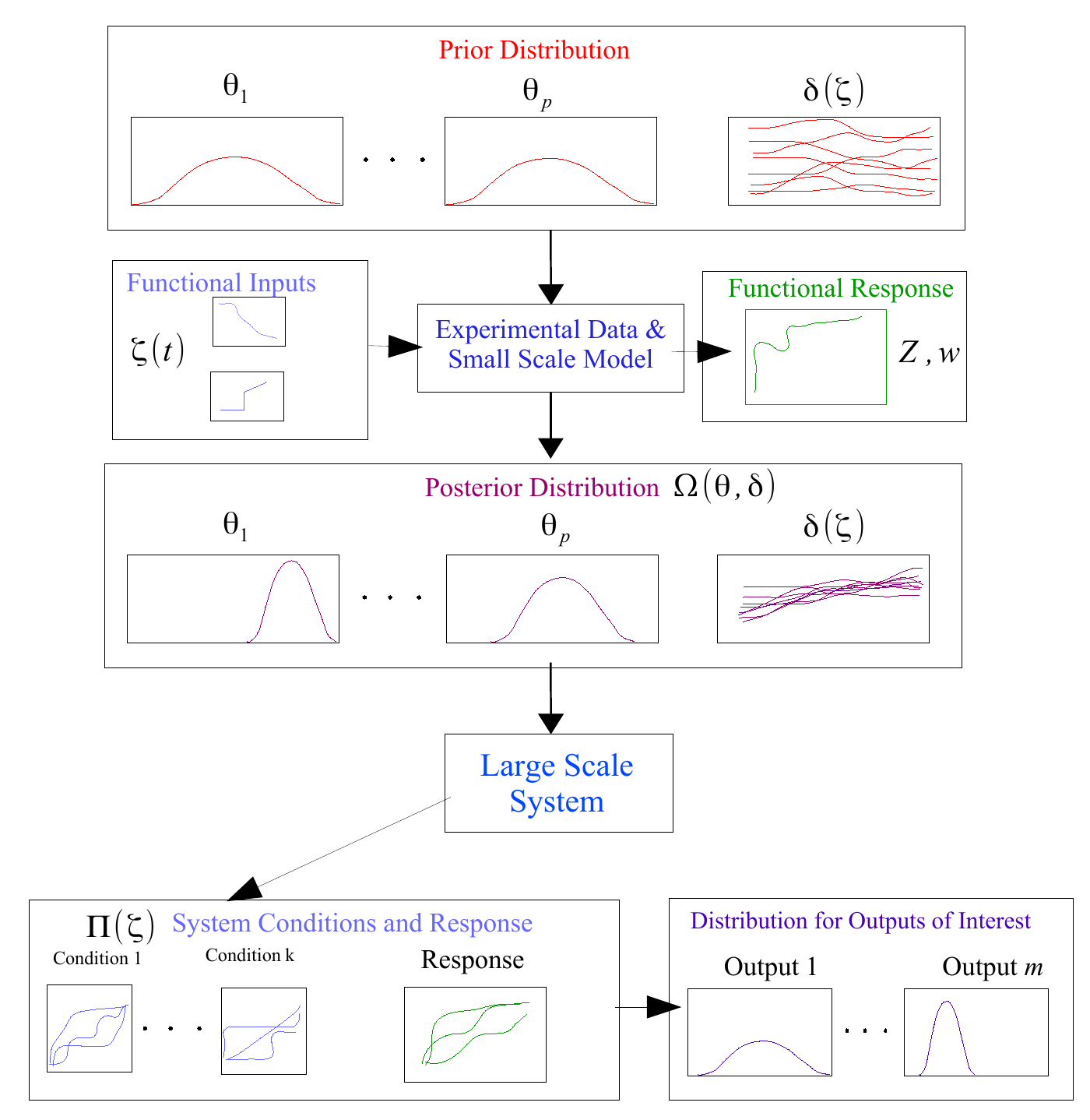} 
%  \figoneAC{Figures/fullupscdiag.pdf}
\vspace{-0.0in}
\label{fig:fullupsc}
\end{center}
\vspace{-.25in}
\end{wrapfigure}   
A road map for the upscaling methodology is provided here.  As shown in the top half of Figure \ref{fig:fullupsc}, the small-scale model is calibrated to experimental data (see Section \ref{sec:calibdyndisc}).  The small-scale model inputs and outputs are functional in nature and experimental data observations are usually only available at a handful of different profiles.  Thus, the functional input space will typically be only sparsely covered by the experimental design.  When the results from the small-scale calibration are upscaled, the system conditions experienced at scale may be far different that those from the small-scale experiments. Thus, it may become important to limit the amount this extrapolation to the extent possible.
%The small-scale model inputs and outputs are functional in nature, and these inputs are very sparse in the functional space of system conditions, which consists of all physically feasible curves.  Experimental data, however, is usually only available at a few functional inputs. When the results from the small-scale model are upscaled, potentially to system conditions far different that the small-scale inputs, there will inevitably will be an issue with extrapolation. 
A Bayesian calibration approach following the Kennedy-O'Hagan framework \citep{kennedy:bcc} is employed to calibrate the small-scale model to data (see Section \ref{sec:calibdyndisc}) with prior distributions (using domain expertise and previous results whenever possible) on the uncertain model parameters and a prior form on the discrepancy (usually a Gaussian Process (GP)).  It is infeasible to use the conventional approach to model discrepancy for forward propagation, in this case, due to the functional inputs and an entirely different fully coupled system at large scale.  Therefore, we introduce a dynamic discrepancy approach in Section \ref{subsec:dyndisc} where the discrepancy is embedded within the rate equations (i.e. Equation (\ref{eq:smallmod})) of the small-scale model.    
 
The end result from calibration is a joint sample-based distribution representing the model parameters and the discrepancy; which is then propagated to the large-scale system, as seen in the bottom half of Figure \ref{fig:fullupsc}.  For each sample, the large-scale system model \textit {simultaneously} solves for the rate equations (e.g. Equations (\ref{eq:largesysx1})-(\ref{eq:largesysx2})) describing the system response and system conditions at each time step, resulting in functional curves for both the system conditions and response.  The collection of these solutions (from the sample-based distribution) provide a set of probabilistic predictions of the system conditions and response, while incorporating uncertainty.  Aggregated output quantities of interest (such as \% CO$_2$ captured by the system) may then be then derived from the response, resulting in probabilistic predictions characterizing the uncertainty of the output quantities of interest.        

%One additional issue that must be discussed here is the issue of extrapolation during upscaling.  The functional input space $\Pi$ ,which consists of the feasible input paths the system can take, can be very large.  We note that the extrapolation uncertainty     
%In this paper, we will not deal with the challenge of reducing the uncertainty due to extrapolation (that is the subject of a different paper and will be discussed briefly in Section \ref{subsec:caveats}).   

%\subsection{Extrapolation with a Functional Input Space} \label{subsec:extrap}

\section{Calibration with Dynamic Discrepancy}\label{sec:calibdyndisc}

%In this section, a framework for model parameter inference from the observations and functional output from a computer model.  In particular, the goal is to calibrate a sorbent model to the synthetic or real TGA data with a functional form as shown in Figure X.  We note that for upscaling, our statistical approach needs to be flexible enough to fit a broad class of functional output, and not solely the output in this particular data set.  We will discuss the data used in more detail in Section \ref{sec:toymodelapp}.    

\subsection{Computer Model Calibration for Functional Data}  \label{subsec:compmodcalib}

Computer model calibration is often used to constrain the computer model to be consistent with experimental data.  Hence, the primary goal of calibration is to find a set of model parameter values that best reproduce the reality of experimental (or field) data.  In the traditional computer model calibration (i.e., inverse problem) setup \citep{kennedy:bcc}, an output $y$ from the physical system is observed (with observational error) at several ($N$) locations of a ``controllable" vector of inputs $\bzeta=[\zeta_1,\dots,\zeta_q]$.  
This physical reality can be approximated by a simulator (i.e., a computer model), $\eta(\bzeta,\btheta)$, where $\btheta=[\theta_1,\dots,\theta_P]$ is a vector of model parameters. If
fixed at an appropriate (unknown) value of $\btheta=\btheta^*$, then  $\eta(\bzeta,\btheta)$ will approximate the reality at $\bzeta$. Our framework also includes a model form discrepancy function $\delta$ that admits the possibility of model bias (from reality). Therefore a general model for the experimental data is
\vspace{-.0in}\beq
\bY_n = \rho \eta(\bzeta_n, \btheta^*) + \delta(\bzeta_n)  + \eps_n,
\label{eq:data_model_0}
\vspace{-.0in}\eeq
$n=1,\dots,N$, where $\bY_n $, $\bzeta_n$, $\eps_n$ are the model output, inputs, and observation error for the $n$-th observation, respectively, and $\rho$ is a regression parameter, representing multiplicative bias.  Due to identifiability issues, $\rho=$1 will be assumed here.
The goal is to estimate $\btheta^*$ (which for ease of notation we refer to as $\btheta$ from now on) and the discrepancy function $\delta$. This is typically done within a Bayesian framework \citep{Higdon04}, where a prior distribution is placed on $\btheta$ and $\delta$ and then updated by conditioning on the experimental data.  A Gaussian process (GP) prior is often selected for the model discrepancy $\bdelta$  \citep{kennedy:bcc}, which we will discuss in more detail in the next section.

For the work in this paper, both the inputs $\bzeta_n$ and the output $\bY_n$ are functional in nature.  For ease of exposition, the domain of the functional input/output space consists of a single variable, time.  This framework can be easily extended when the domain is multivariate.  Hence the inputs are written as $\bzeta_n(\bt)=[\zeta_{n_1}(\bt),\dots,\zeta_{n_q}(\bt)]$, i.e. $\bzeta_{n_i}$ represents the entire input curve for the $i$th input and $n$-th observation.  In practice, $\bt=[t_1 \cdots t_{T^*_n}]$, is discretized over time.  
The output functional curve(s) is expressed as $\bY_n(t), t \in [0,t_{T^*_n}]$, 
%and $\bY$ is the vector obtained by stacking the output over all observations and times.  
The simulator output at model parameters $\btheta$, inputs $\bzeta_n$, and time $t$ can be expressed as $\boldeta(t;\bzeta_n,\btheta)$.  The calibration framework can now be written below:
\vspace{-.0in}\beq 
\bY_n(t) = \boldeta(t;\bzeta_n,\btheta) + \bdelta(t;\bzeta_n)  + \bepsilon_n(t),  t \in [0,T^*_n]
\label{eq:data_model_1}
\vspace{-.0in}\eeq
It should be noted that in general $\boldeta_n(t)$ and $ \bdelta_n(t)$ are dependent on the \textit{entire} input curves $\bzeta_n$, not just the inputs at time $t$.  The observation error, represented by $\bepsilon(t)$, is assumed to be a white noise process with variance $\sigma^2$.  Independence is assumed a priori between $\bdelta$, $\boldeta$, and $\bepsilon$ are assumed. 

\subsection{Dynamic Discrepancy}  \label{subsec:dyndisc}
  
Model discrepancy is considered here in the context of propagation of uncertainty while including the effect of upstream model shortfall in a large-scale system, i.e., not merely for model parameter inference.  For this effort, the discrepancy must incorporate scientific understanding of the deficiencies of the model as well as flexibility to be applicable for a wide range of functional responses and account for extrapolation.  Due to the functional nature of the inputs, discrepancy function $\bdelta(t;\bzeta)$ is a function of the entire input curve $\bzeta$.     
  
The immediate approach would be to construct an appropriate discrepancy function, calibrate the simulator to experimental data, and upscale the joint posterior distribution $\pi(\btheta,\bdelta)$ to the large-scale system.  \cite{Bhatupsc2012} developed such an approach; a small-scale sorbent model was calibrated to data using the framework in Equation (\ref{eq:data_model_1}).  The model discrepancy represents deficiencies in the sorbent model for both equilibrium and kinetic behavior.  The development of this discrepancy assumed a concurrent functional model \citep[see][pp. 280-293]{ramsay2006functional} to deal with the functional nature of the input/output, which was overly simplistic.    
The joint posterior $\pi(\btheta,\bdelta)$ was then upscaled by differentiating the posterior realizations of $\bdelta$ w.r.t time, smoothed, and included in the rate equations of the large-scale process model (see Equations (\ref{eq:largesysx1})-(\ref{eq:largesysx2})).  However, the differentiated posterior discrepancy realizations were very noisy and required heavy smoothing to avoid solver failures.  
% Such a framework would also have to account for the fact that the discrepancy function at any time is a function of the entire input curve.
%Heavy smoothing of these realizations were necessary to avoid solver failures, which raises questions about how representative the smoothed realizations are of the true discrepancy realizations.

We now present a novel alternative approach that alleviates the issues mentioned above.  The main idea is to include the discrepancy $\bdelta$ within the rate equation of the small-scale model in Equation (\ref{eq:smallmod}) as shown below in Equation (\ref{eq:dyn_disc_1}).  We refer to this $\bdelta$ as a {\em dynamic} discrepancy since it allows the dynamic system to change its path depending on the value of $\bdelta$.  Note here that $\frac{\partial x}{\partial t}$  is directly a function of $\bzeta(t)$, and not the entire curve $\bzeta$. Implementing the discrepancy in the rate equation allows us to naturally use a concurrent functional model, and sidestepping the complications with including the entire functional input.        
\vspace{-.0in}\beq
\frac{\partial x}{\partial t}=f_s(x,\btheta,\bzeta(t))+\bdelta(x,\bzeta(t);\bbeta),
\label{eq:dyn_disc_1}
\vspace{-.0in}\eeq

As a GP prior is usually placed on $\bdelta$, and Equation (\ref{eq:dyn_disc_1}) is a stochastic differential equation (SDE).   If $\bdelta$ were a traditional GP then, even for a fixed $\btheta$ and $\bbeta$ (denoting the hyper-parameters of the GP), Equation (\ref{eq:dyn_disc_1}) would result in an SDE. To avoid these complications, we use a BSS-ANOVA GP (which is discussed in more detail in Section \ref{subsec:bssanova}) prior on $\bdelta$, which among other things, has the advantage of admitting a convenient, approximate parametric form, thus containing its entire stochasticity in its parameters $\bbeta$.  In other words, when $\bbeta$ is fixed, $\bdelta$ is entirely specified, and the SDE in Equation (\ref{eq:dyn_disc_1}) becomes an ODE.  In a Bayesian calibration framework, the SDE can be easily integrated within the MCMC routine as follows: at each iteration we propose a set of model and discrepancy parameters, obtain a solution of the state variable(s) from the ODE, evaluate the likelihood, and accept/reject the sample.  This framework provides an avenue to estimate the joint posterior distribution $\pi(\btheta,\bdelta)$ of model parameters and discrepancy, and allows for the forward propagation of uncertainty in the usual sample based manner.
Furthermore, such an approach is generalizable to a broad class of problems.

\subsection{BSS-ANOVA Model}  \label{subsec:bssanova}

The discrepancy function $\bdelta$ is formulated using the BSS-ANOVA GP model \citep{Reich09}; utilizing a covariance function that directly uses the functional components from a functional ANOVA decomposition \citep{Gu02}.  This approach has two very useful properties in the context of upscaling uncertainty: (i) it provides a convenient parametric form which allows to reduce the SDE in Equation (\ref{eq:dyn_disc_1}) to a ODE providing for easy calibration and uncertainty propagation, and (ii) improves computational efficiency substantially, scaling {\em linearly} with the number of data points, as opposed to $\mathcal{O}(N^3)$ for a traditional GP.  In this section we detail the facets of the BSS-ANOVA model necessary to formulate the discrepancy, and further details about the BSS-ANOVA model may be found in the Supplementary Material.  

We again denote the inputs to the computer model as $\bzeta$ with dimension $P$.  The discrepancy may then be represented as:
\vspace{-.15in}\beq
\bdelta(\bzeta) = \beta_0 +\sum_{r=1}^R \bdelta_r(\zeta_r) + \sum_{r<r'}^R \bdelta_{r,r'}(\zeta_r, \zeta_{r'}) + \cdots
\label{eq:delta}
\vspace{-.15in}\eeq
It is assumed that $\beta_0 \sim N(0,\varsigma_0^2)$, and that each main effect functional component is $\bdelta_r \sim GP(0,\varsigma^2_r K_1)$, for some variance parameters $\varsigma^2_r$, $r=0,\dots,R$, and $K_1$ is the BSS-ANOVA covariance function described in \cite{Reich09}.  That is,
\vspace{-.2in}\beq
K_1(u,u') = B_1(u)B_1(u') + B_2(u)B_2(u') - \frac{1}{24}B_{4}(|u-u'|),
\label{eq:BSS_cov}
\vspace{-.2in}\eeq
where $B_l$ is the $l$-th Bernoulli polynomial. The covariance function in (\ref{eq:BSS_cov}) operates on the domain $[0,1]$.  Therefore inputs and parameters must be transformed to $[0,1]$ prior to analysis.
Two-way interaction functions are assumed to be $\bdelta_{r,r'} \sim GP(0,\varsigma^2_{r,r'} K_2)$, where 
\vspace{-.2in}
\beq
K_2((u,v), (u',v')) = K_1(u,u') K_1(v,v').
\label{eq:BSS_2way_cov}
\vspace{-.2in}
\eeq
Three-way or higher order interaction functional components can be defined similarly.  Under this construction, the resulting component GPs are such that they will satisfy the functional ANOVA constraints, e.g.,  $\int \bdelta_r(u)du=0$ and $\int \bdelta_{r,r'}(u,v)du=0$, almost surely.  Any realization from this GP also lies in first order Sobolev space, i.e., absolutely continuous with derivative in $L_2$.

\begin{wrapfigure}{r}{.455\textwidth}
\vspace{-.45in}
\begin{center}
\caption{First eight eigenfunctions from the Karhunen-Lo\'eve expansion for a main effect function from the BSS-ANOVA covariance.}
\vspace{-.00in}
\includegraphics[width=0.435\textwidth]{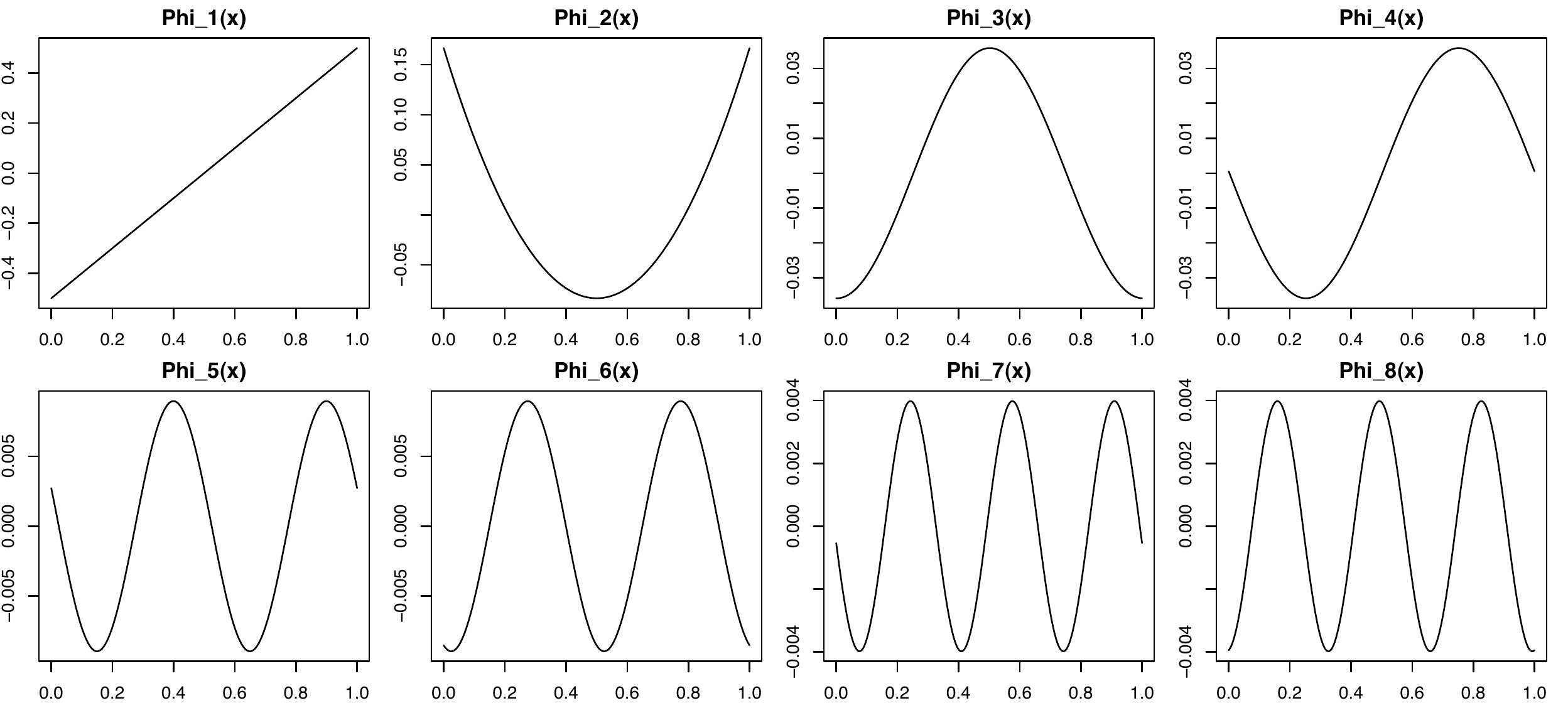} 
\vspace{-0.0in}
\label{fig:KL_basis}
\end{center}
\vspace{-0.5in}
\end{wrapfigure}

It was further demonstrated in \cite{Storlie13onion} that each functional component in (\ref{eq:delta}) can be further written as an orthogonal basis expansion, e.g., 
\vspace{-.1in}\beq
\bdelta_r(\zeta_r) = \sum_{l=1}^\infty \beta_{r,l} \phi_l(\zeta_r), ~~ \beta_{r,l} \stackrel{iid}{\sim} {\cal N}(0,\tau_r^2)
\label{eq:eta_comp}
\vspace{-.1in}\eeq
%\vspace{-.05in}\beq
%\beta_{r,l} \stackrel{iid}{\sim} {\cal N}(0,\tau_r^2)
%\label{eq:alpha_prior}
%\vspace{-.1in}\eeq
The $\phi_l$ terms in the expansion are the eigenfunctions (scaled by the eigenvalues) in the Karhunen-Lo\'eve (KL) expansion (\cite{Berlinet04}, pp.~65-70).  The $\phi_l$ get increasingly higher frequency and have decreasingly less magnitude as depicted in Figure~\ref{fig:KL_basis}, so the expansion in (\ref{eq:eta_comp}) can be truncated at some value $L$.  The choice of $L$ is not critical, as the model will be identical in practice for different $L$ provided it is large enough; our experience suggests that $L \geq 25$ is more than sufficient for most problems.

We use the same decomposition in (\ref{eq:eta_comp}) for two-way and higher interactions as well.  In fact, the $\phi_l$ for two way interactions are simply pairwise products of the corresponding main effect basis functions and similarly for three way and higher interactions.  
In many problems it is sufficient to include only main effects and two-way interactions.  A few preselected three-way interactions selected in consultation with domain experts are included as well.

Hence the overall model in (\ref{eq:delta}) can be written in general as 
\vspace{-.15in}\beq
\bdelta(\bzeta) =  \sum_{j=1}^J \sum_{l=1}^{L^\delta_j} \beta_{j,l} \phi_{j,l}(\bzeta),
\label{eq:delta2}
\vspace{-.2in}\eeq
\vspace{-.0in}\beq
\beta_{j,l} \stackrel{ind}{\sim} {\cal N}(0,\tau_j^2)
\label{eq:beta2}
\vspace{-.0in}\eeq

where (i) $j$ indexes over the $J$ functional components included in the discrepancy realization, and (ii) $l$ indexes over the number of basis functions $L^\delta_j$ used for the $j$-th functional component of the discrepancy representation.  The $\beta_{j,l}$, $\phi_{j,l}$, and $\tau_j$ would correspond to a particular term in the expansion of (\ref{eq:eta_comp}) for the $j$th functional component.  More specific details of the decomposition of the BSS-ANOVA GP into the linear model in (\ref{eq:delta2}) are provided in the Supplementary Material.  

%As mentioned previously, the model for the discrepancy function is also a BSS-ANOVA model, which can be broken down in a similar fashion to 
%\vspace{-.15in}\beq
%\delta(\bx) =  \sum_{k=1}^K \sum_{l=1}^{L^\delta_k} \gamma_{k,l} \psi_{k,l}(\bx),
%\label{eq:delta2}
%\vspace{-.2in}\eeq
%where (i) $\gamma_{k,l} \stackrel{ind}{\sim} {\cal N}(0,\omega_k^2)$, (ii) $k$ indexes over the $K$ functional components included in the discrepancy, and (iii) $l$ indexes over the number of basis functions $L^\delta_k$ used for the $k$-th functional component of the discrepancy. 

%Note that in (\ref{eq:delta2}) the basis functions for each functional component (e.g., $\phi_{j,l}$) do not change with the covariance function parameters ($\beta_j$).  Hence, the KL decomposition of $K_1$ needs to be computed just once on a sufficiently dense grid in one dimension (discussed further in the Supplementary Material), and {\bf no} matrix decomposition is required during the MCMC estimation procedure.  Thus, for estimation purposes, the models for $\bdelta$ is simply a linear model (with $\beta_j$'s as coefficients), and the computational time is $\mathcal{O}(L^\delta N)$, where N is the number of experimental observations.  We discuss this further in Section~\ref{subsec:upscale} and the Supplementary Material.

\subsection {Dynamic Discrepancy for Sorbent Model}  \label{subsubsec:dyndisctoy}

While the additive dynamic discrepancy approach detailed in Section \ref{subsec:dyndisc} is useful in many situations, there were a large number of solver failures due to either a lack of convergence or solutions outside the physical bounds ($x \in [0,0.5]$) for the work in this paper.  The likely cause of this solver instability was the change in sign of the derivative of the rate equation in Equation (\ref{eq:dyn_disc_1}) and/or non-physical solutions when the discrepancy is added.  Further, the manner in which the chemical model could be deficient was deemed by the modelers (as discussed further below) to be more appropriately modeled with a multiplicative discrepancy for the equilibrium and kinetic processes.  This approach, thus, has two advantages; (i) we improve convergence of the solver and guarantee physical solutions when the discrepancy is added, and (ii) the discrepancy here has a clear physical interpretation for sorbent models.    

We first derive the model form discrepancy for the equilibrium process.  We described the rate equation and the equilibrium and kinetic constants $\kappa$ and $k$ for an ideal reaction (which assumes no interaction energy for adsorbates and equivalence among all adsorption sites) in Equation (\ref{eq:kinetic}).  Under the assumption of thermodynamic ideality, $\Delta H$ and $\Delta S$, the enthalpy and entropy respectively, are constant with respect to the thermodynamic state space $\bzeta=\{p,T\}$.  However, when this assumption violated, one frequent way of representing this non-ideality is by allowing $\Delta H$ and $\Delta S$ to depend on the state space $\bzeta$, i.e. Equation (\ref{eq:kinetic}) becomes       
\vspace{-.15in}
\begin{align}
\label{eq:dyndisceqx1}
&\frac{\partial x}{\partial t}=\kappa^K [(1-2x)^2 p-x^2/\kappa^E], ~~~~ \kappa^E=\exp \left(\frac{-\Delta S}{R}\right) \exp \left(\frac{-\Delta H}{RT}\right) \\%=\frac{x^2}{p(1-2x)^2}  \\
\label{eq:dyndisceqx2}
&\kappa^E_{\mbox{\scriptsize{new}}}=\exp \left(\frac{-\Delta S}{R}\right) \exp \left(\frac{-\Delta H}{RT}+\bdelta^E(p,T)\right) \\%=\frac{x^2}{p(1-2x)^2}  \\
\label{eq:dyndisceqx3}
&\kappa^E_{\mbox{\scriptsize{new}}}=\exp \left(\frac{-\Delta S}{R}\right) \exp \left(\frac{-\Delta H}{RT}\right) \exp[{\bdelta^E(p,T)]}=\kappa^E \exp[{\bdelta^E(p,T)]}. %= \frac{x^2}{p(1-2x)^2} 
\vspace{-.15in}
\end{align}
In Equation (\ref{eq:dyndisceqx2}), we have derived an equilibrium constant $\kappa^E_{new}$ that allows for deviations from an ideal reaction, where $\bdelta^E(p,T)$ is a stochastic function which represents this discrepancy.  It is clear from this formulation that for any realization of $\bdelta^E(p,T)$, the chemical state $x \in [0,0.5]$.  The same idea is applicable for the reaction rate constant $\kappa^K_{new}$ as shown in Equation (\ref{eq:dyndisceq2}).  The kinetic rate discrepancy $\delta^k$ is a function of $p$, $T$, and $x$, since the rate of reaction via $k_{\mbox{\scriptsize{new}}}$ will necessarily depend on the current value of $x$ as opposed to the equilibrium constant $\kappa_{\mbox{\scriptsize{new}}}$ which only depends on $p$ and $T$.        
%For the simple small-scale sorbent model in Section \ref{subsec:basicchem}, $\bzeta_E=\{p,T\},\bzeta_K=\{x,p,T\} $
\vspace{-.15in}\beq
\label{eq:dyndisceq2}
\kappa^K_{\mbox{\scriptsize{new}}}=\gamma T \exp \left(\frac{-\Delta H^\ddagger}{RT}+\bdelta^K(x,p,T)\right)=\kappa^K\exp[\bdelta^K(x,p,T)]
\vspace{-.15in}\eeq

The two discrepancy functions, which may be written more generally as $\bdelta^E(\bzeta^E,\bbeta^E)$ and $\bdelta^K(\bzeta^K,\bbeta^K)$ are expressed using a Gaussian Process with a BSS-ANOVA covariance function using Equation (\ref{eq:eta_comp}) as follows.
\vspace{-.15in}\beq
\bdelta^E(\bzeta^E) =  \sum_{j=1}^{J^E} \sum_{l=1}^{L^{\delta^E}_j} \beta^E_{j,l} \phi^E_{j,l}(\bzeta^E),~~~\bdelta^K(\bzeta^K) =  \sum_{j=1}^{J^K} \sum_{l=1}^{L^{\delta^K}_j} \beta^K_{j,l} \phi^K_{j,l}(\bzeta^K).
\label{eq:delta3}
\vspace{-.2in}\eeq
For the simple small-scale sorbent model in this paper, $\bzeta_E=\{p,T\},\bzeta_K=\{x,p,T\}$.  We organize the $\beta$'s as follows, let $\bbeta=[\bbeta^E,\bbeta^K]=[\bbeta^E_1, \cdots, \bbeta^E_{J^E},\bbeta^K_1, \cdots, \bbeta^K_{J^K}]  $, where $\bbeta^E_{j} = [\beta^E_{j,1}, \dots, \beta^E_{j,L^{\delta^E}_j}]^T$ and $\bbeta^K_{j} = [\beta^K_{j,1}, \dots, \beta^K_{j,L^{\delta^K}_j}]^T$ are $L^{\delta^E}(L^{\delta^K})$ vectors of the regression parameters for the $j$-th functional component of the appropriate discrepancy.  We note the BSS-ANOVA formulation includes a constant term $\beta_{0}$, which is completely confounded with the model parameters $\Delta S$ and $\gamma$; thus $\beta_{0}$ is set to zero.  The modified rate equation for the toy sorbent model, with the discrepancies $\bdelta^E$ and $\bdelta^K$ embedded is shown below in Equation (\ref{eq:dyndisceq3}).
\vspace{-.15in}\beq
\label{eq:dyndisceq3}
 \frac{\partial x}{\partial t}=k\exp[\bdelta^K(x,p,T)] ((1-2x)^2 p-x^2/[\kappa \exp[{\bdelta^E(p,T)]}]) 
\vspace{-.15in}\eeq

\subsection{Calibration at Small Scale and Upscaling Uncertainty} \label{subsec:upscale}

We now describe our Bayesian approach to infer a joint posterior probability distribution of the model parameters and discrepancy parameters of the small-scale model.     Let $\by=[\by_1,\dots,\by_N]$ be the experimental observations, $\boldeta(\btheta,\bdelta)=[\eta_1(\btheta,\bdelta),\dots,\eta_N(\btheta,\bdelta)]$ the model output, $\bdelta=[\bdelta^E,\bdelta^K]$ the discrepancy, and $\bepsilon=(\epsilon_1, \cdots, \epsilon_N)^T \sim N(0,\sigma^2 \bI_N)$ the observation error variance.   Combining the results from the previous sections, we express the calibration framework in Equation (\ref{eq:data_model_3}), incorporating the dynamic discrepancy in Section \ref{subsubsec:dyndisctoy}.  Our goal is to estimate $\{\btheta,\bdelta,\sigma^2 \}$ given $\by$, where   
%\vspace{-.15in}\beq
%\bY(t) = \eta(t;\bzeta,\btheta,\bdelta)  + \bepsilon(t),  t \in [0,T]
%\label{eq:data_model_2}
%\vspace{-.15in}\eeq  
\vspace{-.15in}\beq
\by = \boldeta(\btheta,\bdelta;\bzeta) + \bepsilon
\label{eq:data_model_3}
\vspace{-.15in}\eeq  

Prior distributions are required for $\btheta$, $\bdelta$, and $\sigma^2$ to complete the Bayesian model specification.  Priors for some of the model parameters $\btheta$ are derived from \textit{ab initio} quantum chemistry calculations and from prior scientific studies or expert judgement. The model for $\bdelta$ in Equation (\ref{eq:delta2}) require a prior specification for $\tau_j$'s from Equation (\ref{eq:beta2}).   A diffuse Inverse Gamma prior is chosen for both the $\tau_j$'s and $\sigma^2$; the selection of these conjugate priors results in Gibbs updates for these parameters within MCMC procedure.  In Section \ref{sec:toymodelapp}, we will discuss prior selection further, especially for $\btheta$, .  
Letting $\btau=[\tau^E_1, \cdots \tau^E_{J^E},\tau^K_1, \cdots \tau^K_{J^K}]$ the posterior distribution $\pi(\btheta,\bbeta,\btau,\sigma^2 \mid \by) \propto \cL(\by \mid \btheta,\bbeta,\sigma^2)p(\bbeta \mid \btau) p(\btheta) p(\btau) p(\sigma^2)$ is obtained through simulation using Markov Chain Monte Carlo (MCMC). 
The MCMC routine is a hybrid sampling scheme of Gibbs and Metropolis Hastings (MH) updates, where Gibbs updates are viable for $\btau$ and $\sigma^2$ with appropriate conjugate priors and MH updates are necessary for $\btheta$ and $\bbeta$.  
For $\bbeta$, all the coefficients for each main effect and second order interaction components are updated simultaneously, i.e $\bbeta_j$ is using a multivariate normal proposal.   Block updating using joint proposals rather than updating each parameter individually appears to improve mixing and reduces the number of sorbent model evaluations necessary (each proposal requires a model evaluation), which in turn should reduce precious computational time for the MCMC procedure.   

After integrating out the other parameters, we obtain a sample-based distribution of $\pi(\btheta,\bbeta\mid \by)$.  We then draw $n$=200 samples from the posterior distribution $\pi(\btheta,\bbeta\mid \by)$ to upscale to the large scale system, let $\btheta^{(i)},\bbeta^{(i)}$ be the $i$th sample and $\bdelta(\bbeta^{(i)})$ is the discrepancy realization for $i$th sample. We then forward propagate each sample to the large-scale system model, which solves a set of differential equations shown in Equations (\ref{eq:largesyswdiscx1})-(\ref{eq:largesyswdiscx2}), where the dynamic discrepancy $\bdelta(\bbeta)$ has been embedded into the rate expression $\frac{\partial x}{\partial z}$, as in Equations (\ref{eq:dyndisceqx2}) and (\ref{eq:dyndisceq2}).  
 \begin{align}
    \label{eq:largesyswdiscx1}
     &  \frac{\partial x}{\partial z}= f_s(x,\bzeta(z);\btheta,\bbeta) \\
   % \frac{\partial \bzeta_1}{\partial t} &= g_1(x,\bzeta(t);\btheta)\\
  & g_1(x,\bzeta;\btheta)=0\\
    \nonumber
    &~~~~~~~~\vdots \\
     \label{eq:largesyswdiscx2}
    %\frac{\partial \bzeta_q}{\partial t} &= g_q(x,\bzeta(t);\btheta)
  &  g_q(x,\bzeta;\btheta)=0.
\end{align}
The solution to this set of equations for the $i$th sample ($\btheta^{(i)},\bbeta^{(i)}$) is $\bx^{(i)}_l$ and $\bzeta^{(i)}_l$, which are the system response and system conditions (e.g. temperature and partial pressure) respectively.  The final output of the upscaling is a sample based distribution (of $n$ samples),  $[\bzeta^{(i)}_1, \cdots \bzeta^{(i)}_n]$ and $[\bx^{(i)}_1, \cdots \bx^{(i)}_n]$, that may be converted into a distribution of certain quantities of interest.  As discussed earlier, the BSS-ANOVA approach for the discrepancy allows
us to easily calibrate and upscale because its parametric form allows a typical solution to the set of differential equations.  The parametric form also has the effect of improving the computational scalability of the MCMC algorithm to $\mathcal{O}$(n).  Furthermore, provided that the ranges of the inputs used in the small-scale experimentation are sufficiently large, the BSS-ANOVA approach will account for the uncertainty due to extrapolation due to upscaling.  

%As previously mentioned, a major benefit of the BSS-ANOVA approach is the scalability of the resulting MCMC algorithm.  Specifically, the algorithm is $O((J+K)(n+m))$.  This result is formalized in the Supplementary Material, immediately after the full MCMC algorithm details are provided.  Generally, the number of function components for the emulator is $J=O((P+Q)^2$ for a two-way interaction model or $J=O((P+Q)^2+P^2Q)$ for the limited three-way interaction model used in the results presented here.  However, as long as the dimensionality of the model input and parameter spaces are moderate relative to the number of observations and simulator runs, this approach will have a large computational advantage over the existing approaches which are $O((N+M)^3)$.

\section{Application to a Simple Carbon Capture System}\label{sec:toymodelapp}

In this section, we apply our methodology on a simple process model (``bubbling fluidized bed" absorber \citep{lee2012one}) driven by a single small-scale chemical sorbent model described in Section \ref{subsec:basicchem}.   
A post-combustion CO2 capture system consists of two parts; the adsorber which takes up CO$_2$ and the regenerator which lifts CO$_2$ off the sorbent and passes it along for sequestration.  This application focuses on just the bubbling bed adsorber.  The methodology is first illustrated on a truth known example so that its performance can be assessed, and then applied to actual data in Section \ref{subsec:FGdata}.  The small-scale model is calibrated to a ``reality" function, which is actually a more complicated sorbent adsorption process with two chemical reactions with the same inputs and outputs as the single reaction sorbent model.   The calibration results are then upscaled to the process model by propagating uncertainty forward.   This section only provides a brief overview of the ``reality" function and the carbon capture process; more information about this process model is available in the Supplementary Materials.

\subsection{Synthetic Data Generation and Process Model Description} \label{subsec:datadesc1} 
The ``reality" function plus iid Gaussian noise is used as a proxy for the experimental data.  This exercise is used here to illustrate the proposed methodology without data complications and provide a validation of the upscaled results.  This function is based on a two step adsorption process of CO$_2$ by the sorbent according to the reactions below. 
This is a more complicated process than the sorbent model discussed in Section \ref{subsec:basicchem} in that the single reaction in Equation (\ref{eq:carbacid}) is in "reality" two separate reactions.. 
\begin{gather}
  \ce{CO2 + R2NH -> R2NH^{+}-COO-}\\
  \ce{R2NH^{+}-COO- + R2NH <=> R2NCOO^{-}:R2NH2+}
\end{gather}
There are eight ``reality" parameters that need to be specified; $\Delta H_x$, $\Delta
S_x$, $\Delta H_z$, $\Delta S_z$ are enthalpies and entropies for the two reactions, $\Delta H^\ddag_x$, $\Delta H^\ddag_z$, $\gamma_x$, and $\gamma_z$ are the activation energies and pre-exponential factors for the two reaction, and $n_\textrm{v}$ is the number of active amine sites per
unit volume of sorbent; $\btheta^*=[\Delta H_x, \Delta S_x,\Delta H^\ddag_x,\gamma_x,n_\textrm{v},\Delta H_z, \Delta S_z,\Delta H^\ddag_z,\gamma_z]$.  The rate equations of this sorbent model in (\ref{eq:reality}) are solved on a temporal grid, resulting in a functional response $w(t)$ (the sorbent weight gain) with temperature ($T$) and partial pressure ($p$) of CO$_2$ as functional inputs over time (see Figure \ref{fig:funcform} for an example temperature input). 
\begin{align}
 \nonumber
&  \frac{\partial z}{\partial t} = k_z(sp - z/\kappa_z)- k_x(sz - x^2/\kappa_x)\\
 % \kappa_x = \frac{x^2}{zs}\\
 \nonumber
 &\frac{\partial x}{\partial t} =  k_x(sz - x^2/\kappa_x)\\
  \label{eq:reality}
 & s = 1-2x-z, ~~ w = Mn_\textrm{v}x/\rho \\
  \nonumber
 & \kappa_x = \exp{\left(\frac{\Delta
        S_x}{R}\right)}\exp{\left(-\frac{\Delta H_x}{RT}\right)}
  \kappa_z = \exp{\left(\frac{\Delta
        S_z}{R}\right)}\exp{\left(-\frac{\Delta H_z}{RT}\right)}/P\\
  \nonumber
&  k_x = \gamma_x\exp{\left(-\frac{\Delta H_x^\ddag}{RT}\right)}
  k_z = \gamma_z\exp{\left(-\frac{\Delta H_z^\ddag}{RT}\right)}
\end{align}
The sorbent weight gain $w(t)$ is a multiple of the sum of $x$ and $z$, or the fraction of amine sites occupied by carbamic acid and zwitterions respectively, $M$ is the molar weight of CO$_2$, and $\rho$ is the sorbent density, $R$ is the ideal gas constant, and $P$ is the total pressure; the latter four are constant within the model.  The equilibrium constants for the two reactions are $\kappa_x$ and  $\kappa_z$, and the reaction rate constants for the two reactions are $k_x$ and $k_z$.

A few details regarding how the synthetic data used for calibration are now described.  For appropriate functional inputs of partial pressure and temperature along with model parameters $\btheta^*$, the reality model solves for output curve $w(t)$.   The reality model parameters (below) $\btheta^*$, were selected to ensure favorable convergence and system behavior properties when upscaled to the carbon capture process, 
  \vspace{-0.15in}
 \begin{equation*}
 \btheta^*=[-88671, -67.056, 35148, 141.22, 2000, -32055, -87, 53594, 25657].
  \vspace{-0.15in}
 \end{equation*}

\begin{wrapfigure}{r}{.4\textwidth}
\vspace{-.65in}
\begin{center}
\caption{\textit{Temperature input profile for synthetic data.}}
\vspace{-0.20in}
\includegraphics[width=0.45\textwidth]{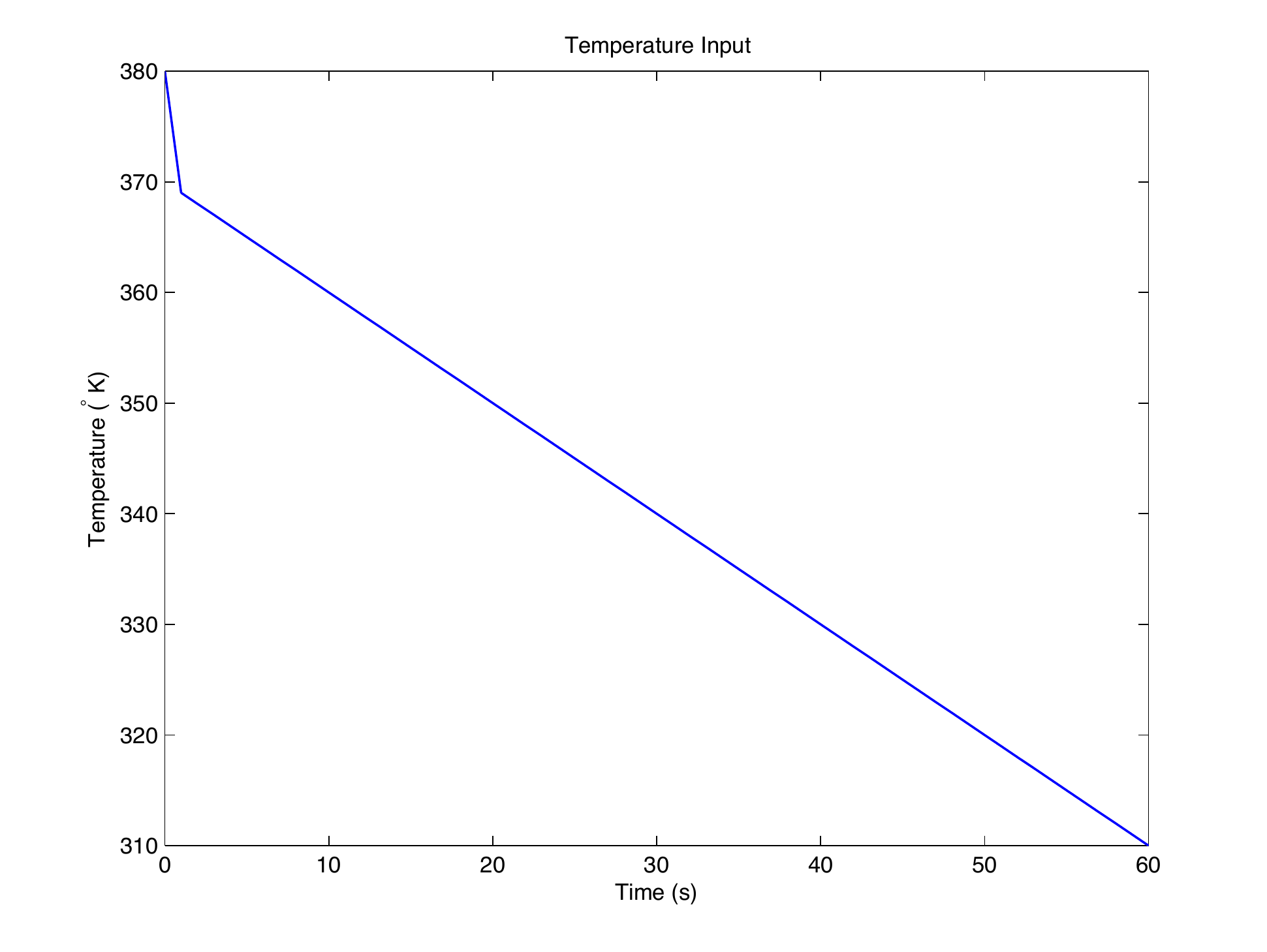} 
%  \figoneAC{Figures/fullupscdiag.pdf}
\vspace{-0.0in}
\label{fig:toytempprof}
\end{center}
\vspace{-.75in}
\end{wrapfigure}  
 
Since the synthetic data selected should be consistent with the behavior of the experimental apparatus, input profiles with constant partial pressure and declining temperature over time are selected.  In particular, five sets of different inputs, with partial pressures 1, 4, 7.5, 10 and 20\%  CO$_2$, which are consistent with partial pressures in an adsorber process.  All five inputs have a temperature profile shown in Figure \ref{fig:toytempprof} where $t$ goes from 1 to 60 seconds.    The starting point ($t$=0) for all four profiles is at a larger pressure and temperature than the rest of the input profile to ensure stability of the model.  A small amount of white noise (with standard deviation of $10^{-4}$) is added to mimic the effect of observation error to the output of the ``reality function".  The output from the four input profiles are stacked to get our synthetic data vector, resulting in a total of N=305 data points.  The small-scale sorbent model will also be run at these same five input profiles. 

%The reality model parameters were selected to satisfy two criteria; one, the reality function resulted in favorable convergence and system behavior properties when upscaled to the carbon capture process.
%and two, to minimize time discretization error at a sufficiently large time step (say $\tau=0.1 s$).  The second criterion is particularly crucial to minimize time discretization error for the carbon capture system, where there are currently methodological limitations regarding the size of the discretization time step.  The parameters for the reality function selected are below.  
The up-scaled physical process, modeled in Aspen Custom Modeler \citep{aspen2011version}, consists of a single device: a one-dimensional, three region ``bubbling fluidized-bed" adsorber with internal heat exchangers, through which the sorbent and a mixture of CO$_2$ and N$_2$ flow in a co-current configuration; both the sorbent and the gas get injected at the bottom and flow upwards. The model predicts the hydrodynamics of the bed and provides axial profiles for all temperature, concentration, and velocities as given in \cite{lee2012one}.  %Lee and Miller, 2012.
Adsorption of CO$_2$ by the sorbent produces heat in accordance with the heat of reaction, which is equivalent to the
adsorption enthalpy $\Delta H$.  This heat is removed by a heat exchanger that runs along the length of the adsorber, thereby
regulating the temperature within the bed.  The full process model is presented in the Supplementary Material, along with a table of fixed process model parameters.
  
% The two-step ``reality" model in Section \ref{subsec:datadesc1} considers both zwitterions and carbamate as part of sorbent components within the process model. Adsorption and desorption reaction kinetics are embedded within the BFB model to obtain ``up-scaled" device-scale process. The steady-state simulation of the process thus obtained contains about 30,000 equations and 130,000 non-zero state variables. 
%The spatial profiles of temperature, CO$_2$ partial pressure and sorbent concentration are extracted from each of these processes in the one of the three hydrodynamic phases; ``emulsion phase" in this case. The spatial profiles are resolved to temporal ones using average of inlet and exit velocities of corresponding phase within each spatial grid. This results in dynamic-profiles analogous to TGA batch data for validation of our approach. 
\subsection{Implementation for Calibration and Upscaling}  \label{subsec:implem1}

The calibration approach discussed in Section 3 is applied to the synthetic data generated in Section \ref{subsec:datadesc1} using Equation \ref{eq:data_model_3}.  This calibration approach requires embedding the discrepancy functions within the model solver as described in Equations (\ref{eq:dyndisceq2}) and (\ref{eq:dyndisceq3}).  The BSS-ANOVA basis functions are not analytical; they are numerically evaluated on a dense grid and represented as continuous functions using linear interpolation. This choice of interpolant enables the function itself as well as its derivatives to be calculated at any point on its domain. A Crank-Nicolson scheme is used to discretize the ordinary differential equation in (\ref{eq:dyndisceq3}), meaning that the system can be solved given the input functions $p$ and $T$ along with the parameter sets $\btheta$ and $\bbeta$ using Newton's method.

As discussed earlier, Bayesian and MCMC methods are deployed to obtain the posterior distributions of $\btheta$, $\bbeta$ and $\sigma^2$.  For three model parameters, $\Delta H$, $\Delta S$ and $n_\textrm{v}$, priors were derived using previous scientific studies; \textit{ab initio} calculations from quantum chemistry calculations were used to derive the following prior, $\Delta H \sim N(-60.84,125 )$ KJ/mol.  These scientific studies were used to derive priors for $\Delta S \sim N(-250, 625)$ J/mol-K, truncated at -200 J/mol-K, and $n_v \sim N(1469,86362)$ mol/m$^3$.  More information about the derivation of the priors may be found in \cite{mebane2013bayesian}.  An empirical approach using a sensitivity study on model convergence were used to obtain priors for the kinetic parameters $\Delta H^\ddagger \sim Unif (-150,-50)$ kJ/mol, and $\gamma \sim Unif (0,5)$.  A joint bivariate normal proposal with correlation parameter $\rho=0.5$ is preferred for block updating $\{\Delta H,\Delta S\}$ as well as $\{\Delta H^\ddag,\gamma\}$; there appears to be a clear relationship between these pairs of model parameters suggested by the rate equations.  

As discussed in Section \ref{subsec:upscale}, conjugate prior distributions are selected for $\btau$ and $\sigma^2$ to ensure Gibbs sampling for these parameters during the MCMC procedure.  A sufficiently diffuse inverse gamma parameters was selected for the elements of $\btau$, specifically $\tau_j \sim IG(0.5,30)$ to allow for adequate flexibility and promote mixing of $\bbeta$.  An inverse gamma prior for the observation error parameter is specified as $\psi \sim IG(1,10^{-8})$.
%M-H sampling is necessary for $\btheta$ and $\bbeta$, while Gibbs sampling is feasible for $\btau$ and $\sigma^2$.  
    
 MCMC was run for 200,000 iterations, allowing for a burn-in of 100,000 samples.  The computer code for the MCMC was implemented in MATLAB using a 2.66 GHz 6-Core Intel Xeon on a Mac Pro desktop with 16GB of RAM.   Obtaining the 200,000 samples using MCMC for N=305 required approximately 190 hours of computer time, the overwhelming majority of which was required to execute the sorbent model.  The sorbent model and ``reality" function were implemented in C++, with a MATLAB executable file; each run of the sorbent model approximately took 6 seconds for each execution.     
%Multiple different starting points were also used to initiate 10 separate chains that all gave very similar results. 
 %To ensure convergence of our MCMC based estimates, we obtained Monte Carlo standard errors for the posterior estimates of parameters computed by consistent batch means \citep{jone:hara:caff:neat:2006,flegal2008mcm}.  The posterior mean estimates of these model parameters had MCMC standard errors below 0.01 times the mean estimates for both the model and discrepancy parameters .  
The carbon capture process model is augmented with the discrepancy functions within Aspen Custom Modeler (ACM) following Equations (\ref{eq:largesyswdiscx1})-(\ref{eq:largesyswdiscx2}) to facilitate the upscaling of the calibration results.  The   Multiple posterior samples are simulated simultaneously by exploiting the multiple parallel ACM run capability offered by CCSI Turbine Gateway developed as part of CCSI project at Lawrence Berkeley National Laboratory (LBNL) facility.  The Gateway utilizes a homotopy-based solver approach leading to high convergence rates for multiple posteriors starting from a common initial state. 
%The software SimSinter, also developed as part of CCSI, is used to configure interface and setup communication between ACM and Turbine Gateway.
%This is being done using a fast C-based external function evaluation (procedure calls) from within the ACM process solver. The resultant steady-state single run require fast computation capability (2.67 GHz 4-Core Intel i7 CPU) and close to maximum memory requirement for a 32-bit machine (~3GB) for storing residual and derivative information for the ``black-box" discrepancy evaluator.   

\subsection{Results} \label{subsec:results}
\begin{figure}[t!]
\begin{center}
\caption{Bivariate marginal posterior distributions for the sorbent model parameters on the off-diagonals, univariate marginal posterior distributions on the diagonals.  The parameters are in displayed in the following order: $\Delta H$, $\Delta S$, $\Delta H^\ddagger$, $\gamma$, $n_v$.}
\vspace{-0in}
\label{fig:posterior_theta}
\includegraphics[width=.95\textwidth]{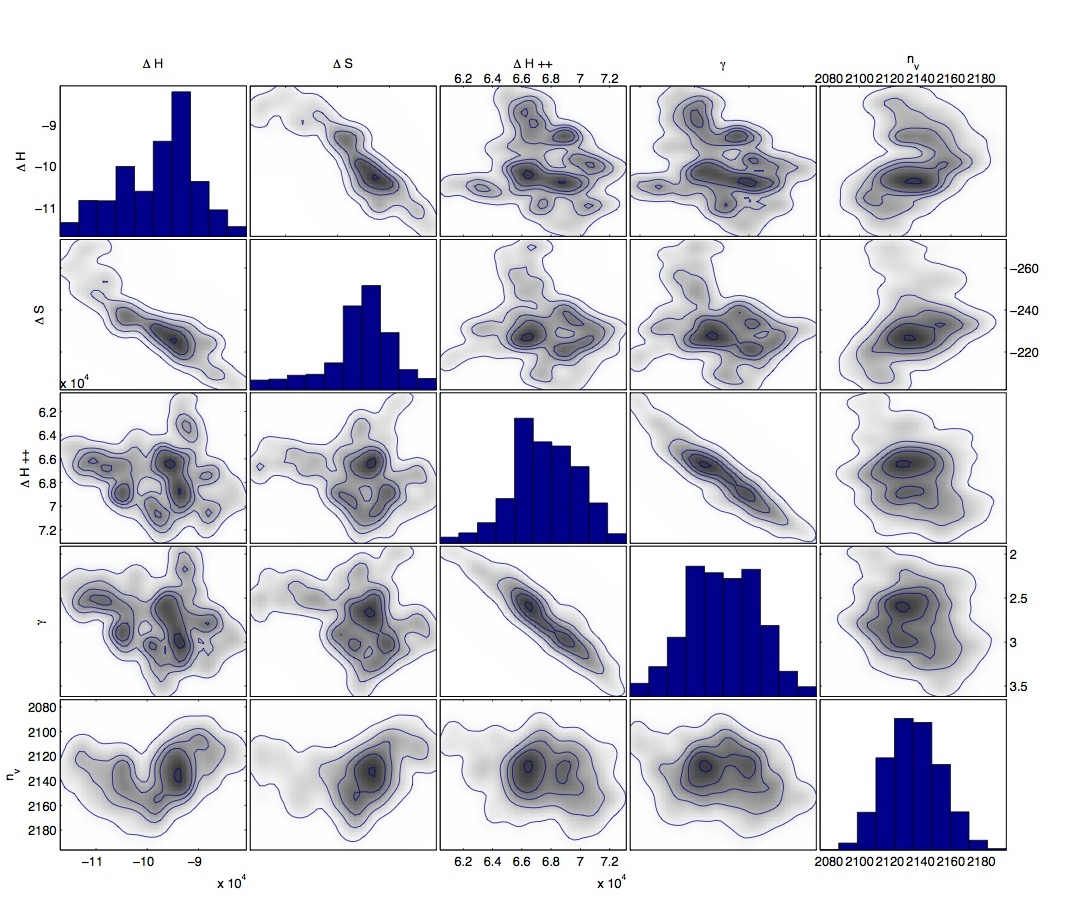}
\end{center}
\vspace{-.45in}
\end{figure}

This section presents results from the calibration of the small-scale sorbent model to the reality function described in Section \ref{sec:calibdyndisc}, and the subsequent upscaling of the calibration results to the simple carbon capture process system.  
The posterior distribution of the model parameters is displayed via bivariate distributions in Figure~\ref{fig:posterior_theta}.  The mean and 95\% credible regions for the sorbent model parameters (see Table \ref{table:HPDmeanCI}) were calculated using the Highest Posterior Density (HPD) method \citep{chen:mcm:2000}.    There are strong correlations between certain pairs of model parameters in the posterior distribution.  The correlation between $\Delta H$ and $\Delta S$ is 0.76, suggesting a very strong relationship between the equilibrium enthalpy and entropy, which seems consistent with the equilibrium analysis of this TGA data set in \cite{mebane2013bayesian}.  In addition, there is a correlation of 0.83 between $\Delta H^\ddagger$ and $\gamma$.      
\begin{table}[h]
\begin{center}
\begin{tabular}{| c | c | c | c | c | }
\hline
Parameter & Mean & 95 \% Lower Bound & 95 \% Upper Bound \\ \hline
$\Delta H$	& -97995 & -112099 & -87252	\\ \hline
$\Delta S$ 	& -231.31 & -260.95 & -210.60 \\ \hline
$\Delta H^\ddagger$	& 67567 & 63633 & 71097 \\ \hline
$\gamma$	& 2.76 & 2.22 & 3.27 \\ \hline
$n_v$ & 2135.2 & 2106.1 & 2185.3 \\ \hline	
%\hline
\end{tabular}
\end{center}
\caption{Posterior mean and 95\% credible regions for sorbent model parameters}
\label{table:HPDmeanCI}
\end{table}%

%It is also clear for some model parameters that the posterior distribution is constricted by physical bounds included in prior distributions and the relationships between parameters.  For the case of amine site density, $n_v$, there is an upper physical boundary of 2351.0 mol/m$^3$ which restricts the range of the posterior distribution.  However, it appears that from further analysis that without this physical boundary, the posterior distribution would place substantial mass on values of $n_v$  larger than the physical bound.  In addition, it appears that the posterior distribution of $\Delta S$ is also constricted by a prior incorporating an upper physical boundary of -200 J/mol-K.   Furthermore, this physical bound along with the strong positive dependence between $\Delta H$ and $\Delta S$ in the model ``pulls" the posterior distribution toward larger values of $\Delta H$. These observations indicate the importance of using scientific prior information, even when there is a large amount of data available, for both parameter inference as well as model development.

\begin{figure}[h!]
\begin{center}
\caption{Posterior fitted plots for sorbent weight gain (\%) for all five CO$_2$ partial pressures.  The data (black open circles), reality (black line), and 30 posterior realizations (green lines), and 95\% credible bands (dashed red lines) are provided for model plus discrepancy predictions.}
\vspace{-0in}
\label{fig:res_toy_preds1}
\includegraphics[width=.95\textwidth]{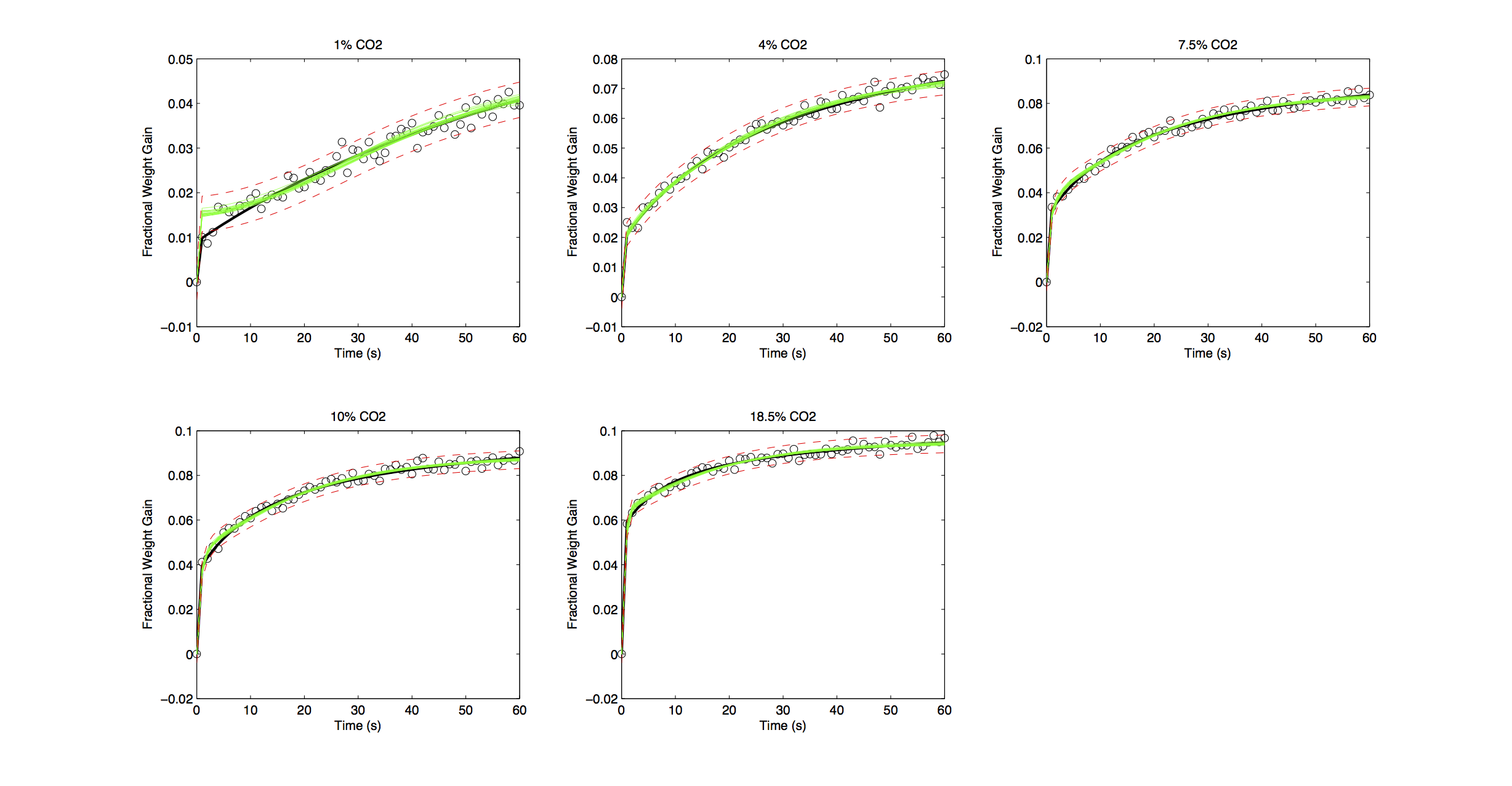}
\end{center}
\vspace{-.45in}
\end{figure}

\begin{figure}[h!]
\begin{center}
\caption{Posterior equilibrium model discrepancy realizations (posterior mean (blue) along with 20 realizations (green) and 95\% credible bands (red)).}
\vspace{-.15in}
\label{fig:res_toy_disc1a}
\includegraphics[width=1.05\textwidth]{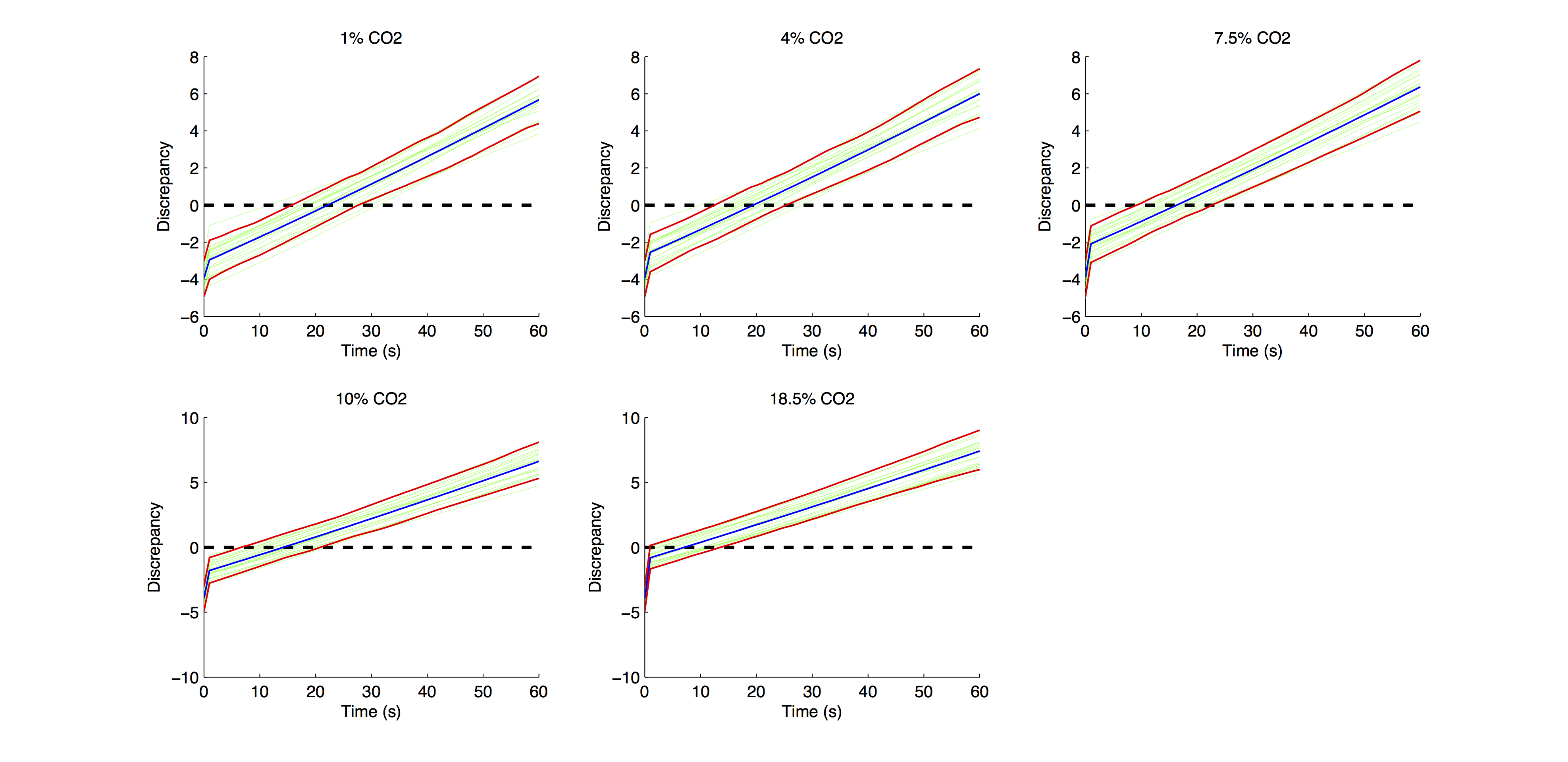}
\end{center}
\vspace{-0.6in}
\end{figure}

\begin{figure}[h!]
\begin{center}
\caption{Posterior kinetic model discrepancy realizations (posterior mean (blue) along with 20 realizations (green) and 95\% credible bands (red)).  The zero line is denoted by a dotted black line.}
\vspace{-.2in}
\label{fig:res_toy_disc1b}
\includegraphics[width=1.05\textwidth]{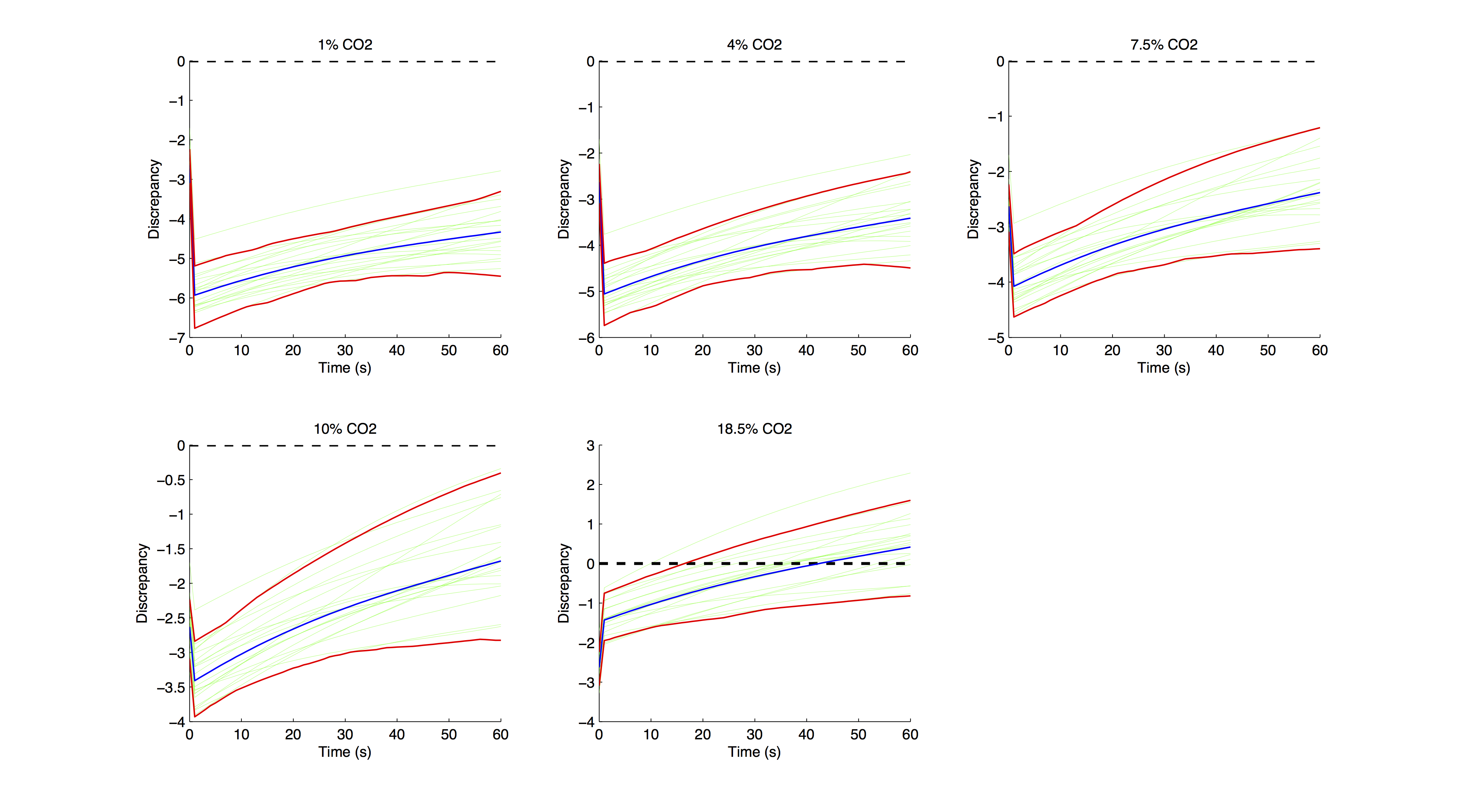}
\end{center}
\vspace{-.3in}
\end{figure}

Posterior predictions and 95\% bounds were computed at 61 time locations for each of the five CO$_2$ composition ratios (1, 4, 7.5, 10, 18.5 \%) for the full calibration approach with the dynamic discrepancy are shown in Figure \ref{fig:res_toy_preds1}.  It is clear that when the dynamic discrepancy is included in making predictions, the data is well represented by the predictions; 30 posterior prediction realizations are shown in green curves.  The discrepancy due to equilibrium and kinetic effects, $\bdelta^E$ and $\bdelta^K$, respectively, are functions of their input curves ($p(t)$, $T(t)$, and $x(t)$).  However, the posterior realizations of $\bdelta^E$ and $\bdelta^K$ given the functional inputs used for calibration (temperature input $T(t)$ in Figure \ref{fig:toytempprof} and constant $p(t)$ corresponding to the CO$_2$ composition ratio) may be expressed as a function of time and suggest a significant non-zero discrepancy which increases over time (see Figures \ref{fig:res_toy_disc1a} and \ref{fig:res_toy_disc1b}).   


\begin{figure}[h!]
\begin{center}
\caption{Results from upscaling posterior distributions obtained from calibrating (to the reality function) small-scale sorbent model to the large-scale carbon capture system.  Distribution of carbon capture rate (left) with reality capture rate (red dot), along with distributions of system conditions; temperature (middle) and partial pressure (right).  Reality function system conditions (black line) and realizations of system conditions from model (red lines) are shown.}
\vspace{-.0in}
\label{fig:res_toy_upsc1}
\includegraphics[width=1.05\textwidth]{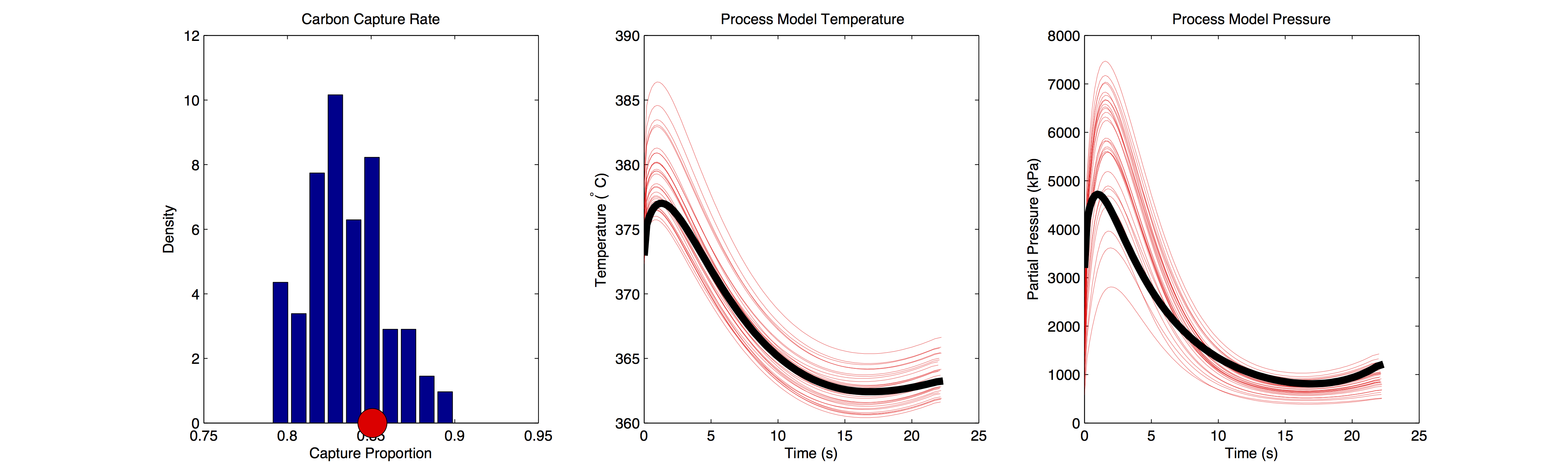}
\end{center}
\vspace{-.1in}
\end{figure}

The results of upscaling the uncertainty from the model parameters and discrepancy to the carbon capture system, hence obtaining a distributions of the capture fraction and the system input conditions are presented below.  200 posterior samples are propagated from the joint distribution of the model parameters and discrepancy to the carbon capture process, as well as implementing the reality function into the carbon capture process.  The 95\% credible region for the carbon capture fraction is between 0.79 and 0.89, which clearly covers the reality capture rate of 0.85 (see Figure \ref{fig:res_toy_upsc1}).  The carbon capture system input conditions for both temperature and partial pressure are also displayed in Figure \ref{fig:res_toy_upsc1}; the distribution of the upscaled model results for both inputs cover the reality input conditions.  There is substantial uncertainty in the upscaled input conditions as well as the capture fraction results; however this is in large part due to the extrapolation due to limited knowledge of the input conditions in the data.  Fortunately, this does not translate into large uncertainty in the primary quantity of interest (i.e., capture fraction).         

%\begin{wrapfigure}{l}{.5\textwidth}
%\vspace{-.45in}
%\begin{center}
%\caption{Process model for simple carbon capture system.}
%\vspace{-.00in}
%\includegraphics[width=0.5\textwidth]{Figures/procmod2a.pdf} 
%%  \figoneAC{Figures/fullupscdiag.pdf}
%\vspace{-0.0in}
%\label{fig:procmod}
%\end{center}
%\vspace{-.25in}
%\end{wrapfigure} 

\subsection{Analysis and Results for Application to Thermogravimetric (TGA) Data}\label{subsec:FGdata}

\begin{figure}[h!]
\begin{center}
\caption{Left: entire output from the TGA experiment at 18.5 \% CO$_2$ (blue line) and temperature (green line), red lines denote boundaries of the selected snippet.  Right: Data from the selected snippet, note that the response shows substantial change over time.}
\vspace{-.0in}
\label{fig:funcform1}
\includegraphics[width=1.05\textwidth]{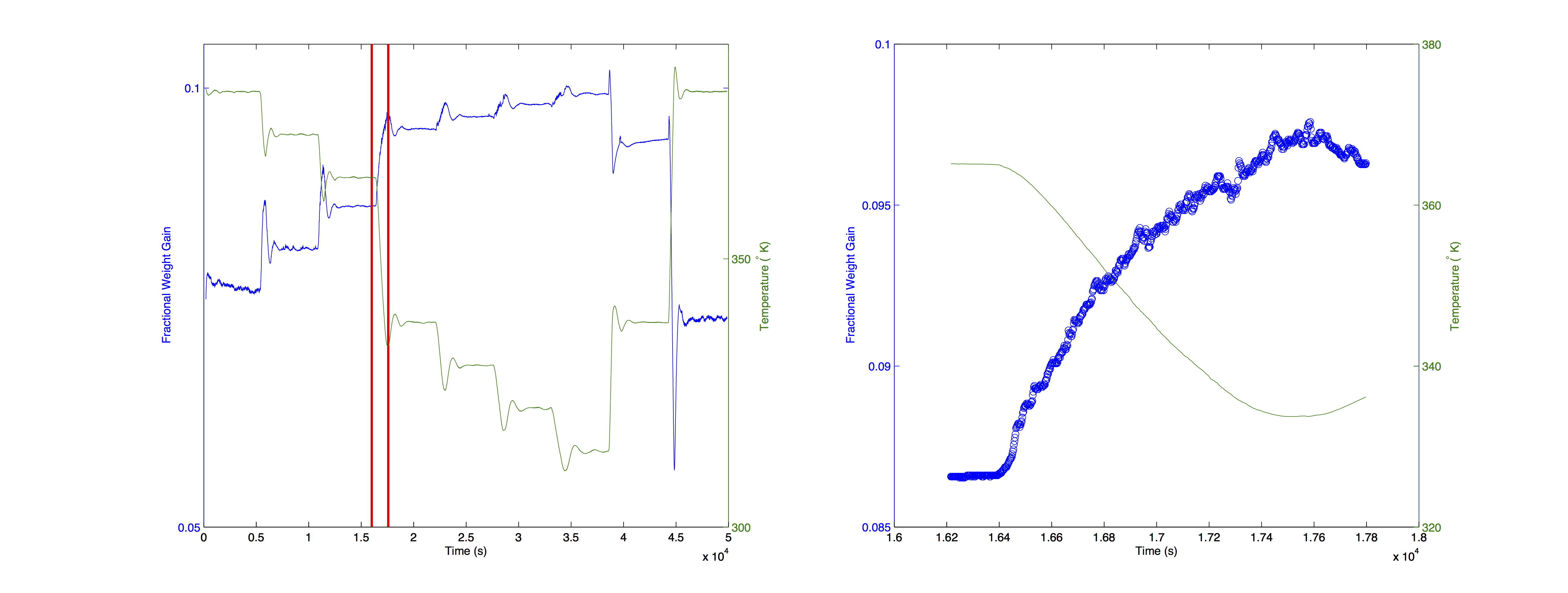}
\end{center}
\vspace{-.1in}
\end{figure}

In this section, the calibration and upscaling with dynamic discrepancy approach in this paper is applied to real experimental data obtained from Thermogravimetric analysis (TGA) to predict the capture fraction of CO$_2$ in a large-scale process model.  TGA was used to obtain the experimental data to constrain the model; these experiments were conducted at the following ratios, 4, 7.5, 10, 18.5, and 100 \%  CO$_2$ (versus N$_2$), each have an associated partial pressure which is constant over time.  The inputs to the TGA experiment consist of that constant partial pressure and a functional temperature curve.   More information about the mechanics about the TGA experiment may be found in \cite{mebane2013bayesian}.  Since the entire TGA response curve for any particular experiment has a domain of up to 200,000 seconds and consists largely of plateau regions (see Figure \ref{fig:funcform1}) which yield little information about the kinetics; snippets of the response to large temperature changes are analyzed instead.  In particular we have carefully chosen 12 snippets from the TGA experiments with partial pressure less than 100 \%.

\begin{figure}[h!]
\begin{center}
\caption{Posterior fitted plots for snippets of TGA data.  The data (black open circles), reality (black line), and 30 posterior realizations (green lines), and 95\% credible bands (dashed red lines) are provided for model plus discrepancy predictions.}
\vspace{-0in}
\label{fig:res_full_preds1}
\includegraphics[width=.95\textwidth]{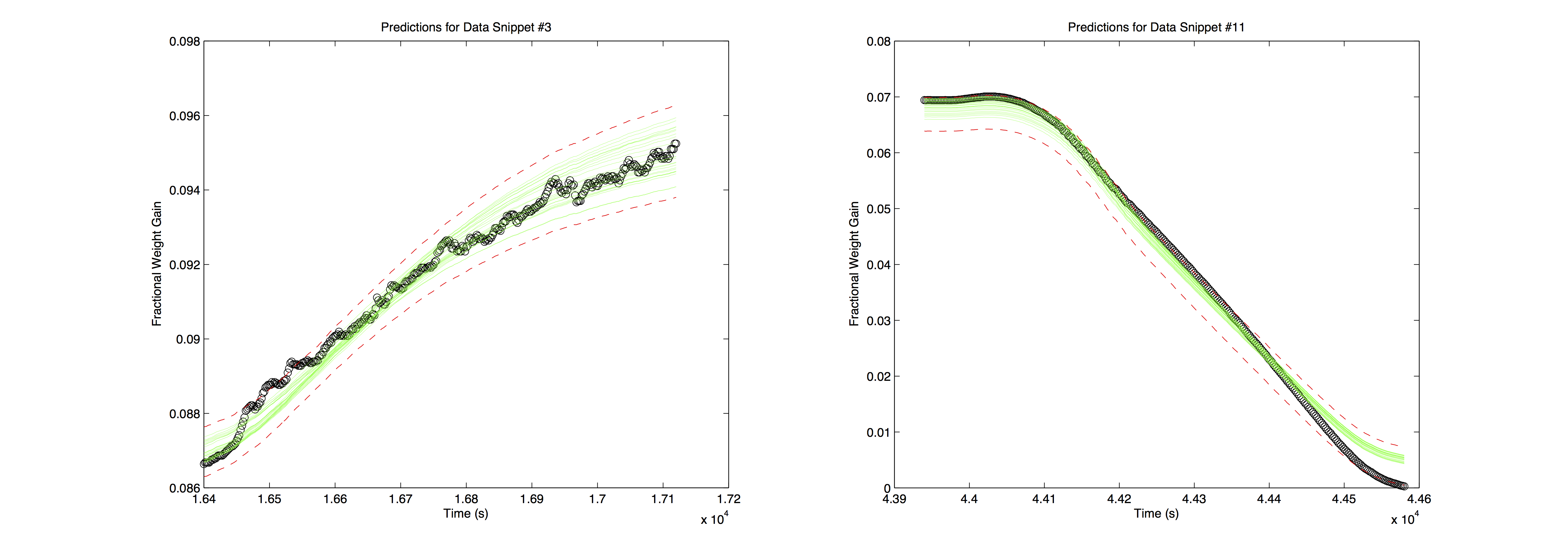}
\end{center}
\vspace{-.45in}
\end{figure}

%Posterior predictions and 95\% bounds were computed at 61 time locations for each of the five CO$_2$ composition ratios (1, 4, 7.5, 10, 18.5 \%) for the full calibration approach with the dynamic discrepancy are shown in Figure \ref{fig:res_toy_preds1}.  It is clear that when the dynamic discrepancy is included in making predictions, the data is well represented by the predictions; 30 posterior prediction realizations are shown in green curves.  The posterior realizations of the discrepancy due to equilibrium and kinetic effects, $\bdelta^E$ and $\bdelta^K$ respectively suggest a significant non-zero discrepancy which increases over time (see Figures \ref{fig:res_toy_disc1a} and \ref{fig:res_toy_disc1b}).   
%%

\begin{figure}[h!]
\begin{center}
\caption{Results from upscaling posterior distributions obtained from calibrating (to the TGA data) small-scale sorbent model to the large-scale carbon capture system.  Distribution of carbon capture rate (left) along with distributions of system conditions; temperature (middle) and partial pressure (right).  Realizations of system conditions from model (red lines) are shown.}
\vspace{-.0in}
\label{fig:res_real_upsc1}
\includegraphics[width=1.05\textwidth]{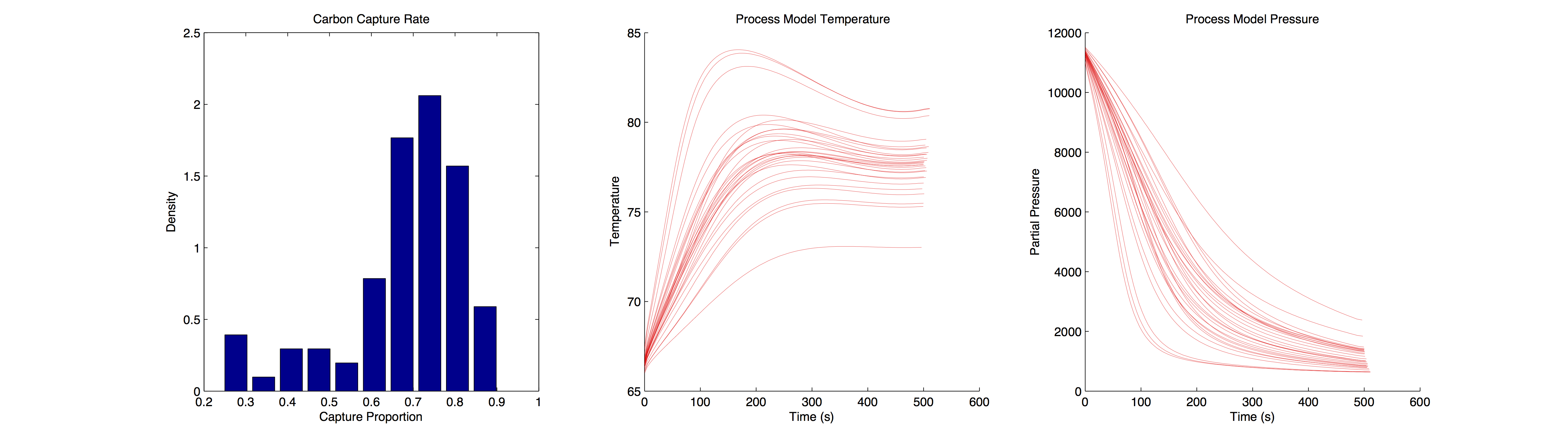}
\end{center}
\vspace{-.1in}
\end{figure}

Posterior predictions and 95\% bounds were computed at the 12 previously selected snippets from the TGA experimental data.  When the dynamic discrepancy is included in the framework, the data appears to be largely covered by the 95\% intervals (see Figure \ref{fig:res_full_preds1} for two snippets).  The results of upscaling the uncertainty from the model parameters and discrepancy to the carbon capture system, hence obtaining a distributions of the capture fraction and the system input conditions are presented here.  One hundred posterior samples are propagated from the joint distribution of the model parameters and discrepancy to the carbon capture process; 18 of these samples failed to converge during upscaling, resulting in 82 samples.  The mean carbon capture fraction is 0.68, and the 95\% credible region for the carbon capture fraction is between 0.32 and 0.85, (see Figure \ref{fig:res_real_upsc1}).  The carbon capture system input conditions for both temperature and partial pressure are also described in Figure \ref{fig:res_real_upsc1}.  Certain fixed system design features (such as flow rate) for the carbon capture are not optimized here, and hence the results of carbon capture fraction is valid only under these conditions.  A full system design optimization under the presence of uncertainty is a natural next step, but beyond the scope of this paper.

%There is substantial uncertainty in the upscaled input conditions as well as the capture fraction results; however this is in large part due to the extrapolation due to limited knowledge of the input conditions in the data.  Fortunately, this does not translate into large uncertainty in the primary quantity of interest (i.e., capture fraction).    

\section{Discussion}\label{sec:discussion}

\subsection{Summary}\label{subsec:summary}    

A novel approach for upscaling uncertainty for a multi-model system with functional inputs has been presented and demonstrated here.  The small-scale model is calibrated to data and then the results of the calibration, a joint distribution of both parameter and dynamic model form discrepancy, are propagated to the large-scale system.  The model form uncertainty is represented by a dynamic discrepancy embedded with in the rate equation(s) in the small-scale model, and has a clear physical understanding of the deficiency of the model.  
The dynamic discrepancy is modeled by a Bayesian Smoothing Splines ANOVA framework, which provides a convenient form for calibration and upscaling, accounts for extrapolation uncertainty and has linear complexity with the number of data points.  The methodology was demonstrated for a simple carbon capture system driven by a small-scale chemical sorbent model. 

\subsection{Caveats and Future Directions}\label{subsec:caveats}

%One potential issue we may encounter is the varying time discretization steps between the computer models in the small-scale model and the large-scale system.  In many cases, including this carbon capture system, there may be limits on the size of the time discretization step due to computational feasibility.  This could require different time steps between the scales, as large time steps for the solvers of some small-scale models may result in numerical instability and inaccurate solutions. We do want to make it clear that there are no physical processes in this systems that are resolved only at very low time resolutions.  The argument has been made that solutions from differential equations from complex systems should be considered as stochastic due to the inconsistency in solutions at different time discretization steps \citep{chkrebtii2013bayesian}.

One major issue with our approach is the computational time which is required to compute the sorbent model, which is the bottleneck in obtaining the posterior distribution for the model parameters and discrepancy.  Each evaluation of the sorbent model requires approximately 6 seconds, and each iteration of MCMC requires 15 model evaluations, requiring one week to obtain the full posterior.  Both the speed of the sorbent model and the number of model evaluations are a function of the number of inputs and discrepancy coefficients needed.  When more complicated models are involved with multiple outputs, the number of differential equations to be solved increases, which in turn increases the number of inputs and discrepancy functions and coefficients.  The end result is that the computational time required for this approach in the current implementation may be prohibitive.    
The computational efficiency of this approach may be improved in three ways.  First, sorbent models can be implemented using graphics processing units (GPU); GPUs often speed up computer models manyfold.  Also, adaptive proposals to reduce the number of MCMC iterations required to reach convergence may be investigated.  Finally, dynamic emulation may be considered as a surrogate for the sorbent model, which would reduce the computational costs of the model evaluations.   
    
%\begin{wrapfigure}{l}{.3\textwidth}
%\vspace{-.55in}
%\begin{center}
%\caption{Iterative method to reduce extrapolation.}
%\vspace{-.00in}
%\includegraphics[width=0.3\textwidth]{Figures/iterative1.png} 
%%  \figoneAC{Figures/fullupscdiag.pdf}
%\vspace{-0.0in}
%\label{fig:iterative1}
%\end{center}
%\vspace{-.80in}
%\end{wrapfigure} 

Extrapolation over the input function space during upscaling in a multi-model context is a difficult challenge.  Our proposed approach demonstrates the inclusion of that uncertainty in the conditions and quantities of interest of the large scale model.  However, this extrapolation could lead to uncertainties that are unacceptably large.  One possible approach to ameliorating this uncertainty due to extrapolation is using a three-step iterative method, which is analogous to the upscaling framework suggested in this paper.  
Data could be gathered at a set of inputs drawn from the sample-based posterior distribution of system conditions after the initial calibration was performed.  This could then be followed by a second calibration of small-scale results using data collected at inputs resembling these upscaled system conditions.  The second calibration results could then be used to once again upscale to the large-scale system.  The entire process could be repeated, each time producing more relevant small scale experiments.

%We are currently working on an iterative approach that would reduce the uncertainty due to extrapolation using a three step method shown in Figure \ref{fig:iterative1}, which is analogous to our upscaling framework suggested in this paper.  First, we gather data at a set of input conditions (e.g. pressure and temperature curves) from the sample-based distribution of system conditions (see Figure \ref{fig:res_toy_upsc1}).  Second, we calibrate the small-scale model to the data we have obtained from the first step, and third, we upscale the calibration results to the large-scale system.  
%We are currently working on convergence properties for this iterative approach to reducing uncertainty due to extrapolation, and conjecture that convergence to the true response will be achieved under a broad set of conditions.     
% 
%
%Also current/future directions (work on iterative approach as well as dynamic emulation).

\section* {Acknowledgements}
 This work was partially supported by the Carbon Capture Simulation Initiative.  The authors also thank Joel Kress for \textit{ab initio} calculations of quantum chemistry data used for priors of model parameters, Andrew Lee and Dan Fauth for helping to develop the sorbent model and providing us the TGA data, Brenda Ng and others at LLNL and LBNL for helping us operate the turbine gateway to set up and run the process model, and Dan Fauth and Mac Gray for relevant figures.  
This report was prepared as an account of work sponsored by an agency of the United States Government. Neither the United States Government nor any agency thereof, nor any of their employees, makes any warranty, express or implied, or assumes any legal liability or responsibility for the accuracy, completeness, or usefulness of any information, apparatus, product, or process disclosed, or represents that its use would not infringe privately owned rights. Reference herein to any specific commercial product, process, or service by trade name, trademark,  manufacturer, or otherwise does not necessarily constitute or imply its endorsement, recommendation, or favoring by the United States Government or any agency thereof. The views and opinions of authors expressed herein do not necessarily state or reflect those of the United States Government or any agency thereof.

\newpage

\maketitle

\newpage

\bibliographystyle{apalike} 
\bibliography{DynamicDiscpaper.v4.arxiv}

\newpage

%Figures
%\begin{figure}
%\begin{center}
%  \figoneAC{Figures/fullupscdiag.pdf}
%\end{center}
%\caption{Overview of the upscaling process.}
%\label{fig:fullupsc}
%\end{figure}

%\begin{figure}
%\begin{center}
%  \figoneAD{Figures/tempfunc.eps}
%\end{center}
%\caption{TGA output for an experiment at 18.5\% CO$_2$ in blue, temperature profile in green.  }
%%%2005: {\color{green} {$\blacktriangle$}}: $<$July 6, {\color{Brown} $\blacklozenge$}: July 6 to July 12, {\color{Orange} $\bullet$}:July 13 to July 18, {\color{red} $\blacksquare$}: after July 20.}
%\label{fig:profile} 
%\end{figure}

\end{document}